\documentclass[12pt]{article}
\usepackage{setspace}
\usepackage{times}
\usepackage{amsfonts}
\usepackage{enumitem}
\usepackage{soul}
\usepackage{float}
\usepackage{latexsym}
\usepackage{nicefrac}
\usepackage{booktabs}
\usepackage[ruled]{algorithm2e}

\SetAlFnt{\small}
\SetAlCapFnt{\small}
\SetAlCapNameFnt{\small}
\SetAlCapHSkip{0pt}
\IncMargin{-\parindent}
\usepackage{mathtools}
\usepackage[
  bookmarks=true,
  bookmarksnumbered=true,
  bookmarksopen=true,
  pdfborder={0 0 0},
  breaklinks=true,
  colorlinks=true,
  linkcolor=black,
  citecolor=black,
  filecolor=black,
  urlcolor=black,
]{hyperref}
\usepackage{xcolor}
\usepackage[margin=1.5in]{geometry}
\usepackage{amssymb}
\usepackage{amsmath}
\usepackage[square]{natbib}
\setcitestyle{numbers}
\usepackage{amsthm}
\usepackage[capitalize]{cleveref}
\usepackage{bbm}
\usepackage{tikz}
\usetikzlibrary{decorations.pathreplacing,angles,quotes}
\theoremstyle{plain}
\newtheorem{theorem}{Theorem}[section]
\newtheorem{corollary}[theorem]{Corollary}
\newtheorem{lemma}[theorem]{Lemma}
\newtheorem{proposition}[theorem]{Proposition}

\newtheorem*{goal*}{Goal}

\theoremstyle{definition}
\newtheorem{definition}[theorem]{Definition}

\newtheorem{example}[theorem]{Example}
\DeclarePairedDelimiter\parentheses{(}{)}
\DeclarePairedDelimiter\braces{\{}{\}}
\DeclarePairedDelimiter\brackets{[}{]}
\DeclarePairedDelimiter\absolute{|}{|}

\title{A Random Dictator Is All You Need}
\hypersetup{
  pdfauthor      = {Itai Arieli <iarieli@technion.ac.il>, Yakov Babichenko <yakovbab@technion.ac.il>, Inbal Talgam-Cohen <inbaltalgam@gmail.com>, Konstantin Zabarnyi <konstzab@gmail.com>},
  pdftitle       = {A Random Dictator Is All You Need}
}
\date{November 15, 2023}
\author{Itai Arieli\thanks{Technion -- Israel Institute of Technology | \emph{E-mail}: \href{mailto:iarieli@technion.ac.il}{iarieli@technion.ac.il}.} \and Yakov Babichenko\thanks{Technion -- Israel Institute of Technology | \emph{E-mail}: \href{mailto:yakovbab@technion.ac.il}{yakovbab@technion.ac.il}.} 
\and 
Inbal Talgam-Cohen\thanks{Tel Aviv University and Technion -- Israel Institute of Technology | \emph{E-mail}: \href{mailto:inbaltalgam@gmail.com}{inbaltalgam@gmail.com}.} 
\and Konstantin Zabarnyi\thanks{Technion -- Israel Institute of Technology | \emph{E-mail}: \href{mailto:konstzab@gmail.com}{konstzab@gmail.com}.}}

\newcommand{\reg}{{\mathrm{Reg}}}

\newcommand{\supp}{{\mathrm{supp}}}

\newcommand{\cav}{{\mathrm{cav}}}

\newcommand{\minmax}{{\mathrm{Minmax}}}
\newcommand{\appr}{{\mathrm{Appr}}}
\newcommand{\vex}{{\mathrm{vex}}}
\begin{document}
\maketitle
\begin{abstract}
We study information aggregation with a decision maker aggregating binary recommendations from symmetric agents. Each agent's recommendation depends on her private information about a hidden state. While the decision maker knows the prior distribution over states and the marginal distribution of each agent’s recommendation, the recommendations are adversarially-correlated. The decision maker's goal is choosing a robustly-optimal aggregation rule.

We prove that for a large number of agents, for the three standard robustness paradigms -- minimax, regret and approximation ratio -- the unique optimal aggregation rule is \emph{random dictator}. We further characterize the minimal regret for any agents' number through concavification.
\end{abstract}
\section{Introduction}
\label{sec:intro}
Every day, we receive conflicting forecasts regarding such issues as investment, health, or even the weather. Based on these forecasts, we must make decisions like whether to buy or sell, to operate or not to operate, to go out or stay home. In most cases, when basing our decisions on the predictions we have, we are not aware of how the information obtained from different sources is correlated. A fundamental challenge presented recently by~\cite{de2021robust} is -- how to aggregate the forecasts and reach decisions despite this knowledge gap.

The practice of \emph{information aggregation} is a common technique used to improve the accuracy of prediction models. Combining forecasts obtained from multiple sources can help account for the inherent uncertainty and variability present in any forecasting task. There is a variety of methods for combining forecasts, including simple averaging~\cite{clemen1986combining}, median averaging~\cite{jose2008simple}, or weighting each forecast based on its past performance~\cite{diebold2002comparing}. Forecast aggregation has been extensively studied in many fields, such as finance~\cite{banbura2013now}, marketing~\cite{armstrong2001principles} and meteorology~\cite{raftery2005using}.

Given the multitude of approaches, to produce reliable and meaningful results one must carefully consider the underlying assumptions and limitations of each aggregation method. A common way to evaluate a solution to a decision problem under uncertainty in general, and to the problem of information aggregation under uncertainty in particular, is via the methodology of \emph{robustness}. That is, the performance of a decision rule is evaluated according to the worst-case (a.k.a.~adversarial)~scenario.

Within the robustness methodology, there are two leading paradigms for quantifying the quality of a decision under uncertainty: the \emph{minimax} approach and the \emph{regret} approach. The minimax approach~\cite{Wald50,gilboa2004maxmin} involves choosing the option that minimizes the maximum possible loss. This approach is based on the idea that the decision maker wants to optimize the worst-case outcome, and thus applying it does not require a benchmark. Alternatively, the regret approach~\cite{savage1951theory} sets an ambitious benchmark -- the loss from a decision made without the burden of uncertainty. It then aims to minimize the difference (known as the regret) between the performance of the optimal decisions with and without uncertainty -- that is, to mimic as closely as possible the optimal informed~decision.

In a typical scenario, the minimax course of action differs from the regret-minimizing course of action. For example, the celebrated result of~\citet{carroll2017robustness}, establishing the minimax-optimality of selling items separately in a combinatorial auction setting, does not hold for regret minimization (see Appendix~\ref{ap:vs}). Such conflicting recommendations weaken the predictive, as well as prescriptive, power of the robustness paradigm.

The methodology of robustness, and specifically -- the minimax approach, is applied by~\citet{de2021robust} to an information aggregation setting. In this paper, we follow their lead by applying the complementary approach of regret to a basic information aggregation setting with symmetric agents. Surprisingly, we show that the simple aggregation rule of randomly choosing an agent to make the decision, known as the \emph{random dictator} rule~\cite{gibbard1977manipulation, sen2011gibbard}, is optimal for both robustness approaches -- minimax and regret -- for a wide range of instances. This result can be viewed as strengthening the message of~\cite{de2021robust}, but it requires different proof techniques. Moreover, the same rule turns out to be asymptotically (in the number of agents) optimal for the \emph{approximation ratio} robustness approach, which is common in computer science~\cite{vazirani2001approximation} and some game theory applications~\cite{RoughgardenTC19}. Our main takeaway is that in our binary decision setting, the random dictator rule is asymptotically universally optimal with respect to all standard robustness concepts. This asymptotic result holds despite the significant conceptual difference between the robustness paradigms, as our results for a small number of agents demonstrate.
\subsection{Our Information Aggregation Setting}
\paragraph{A leading example.}
We now demonstrate our basic information aggregation setting; many of the simplifying assumptions can later be relaxed (see Section~\ref{sec:extensions}). Consider a \emph{decision maker (DM)} facing the decision of whether or not to implement a project. An unknown binary \emph{state of nature} (or \emph{state} for short) indicates whether the project is beneficial or detrimental in terms of utility. As a concrete example, consider the decision of whether to implement containment measures during a pandemic in a certain region (the project). This project's utility depends on whether the illness is on the rise in that region or not (the state of nature). We consider an identical-interest scenario in which the DM and all forecasting agents agree on the utility of the project in both the beneficial and detrimental states.

Each agent is privately exposed to some information about the state. For instance, this can be the number of illness cases in an agent's social network.\footnote{During the height of the COVID-19 pandemic, such information was found to be extremely useful in forecasting the spread of illness in the short term -- see~\cite{Varian20}.} Based on this information, the agents update the probability with which they believe the beneficial state to hold. The agents are symmetric in the sense that all agents who witness a rise (respectively, decline) in illness around them hold the same belief regarding this probability. Formally, agents have identical marginal distributions of posterior beliefs, with correlations unknown to the DM. Each agent {anonymously fills in} a survey reporting her true recommendation as to whether the project should be implemented.\footnote{We assume that agents' recommendations are truthful -- an agent recommends implementation of the project if and only if according to her private information, the expected utility of the project is non-negative. This assumption is natural for many surveys, e.g., if the agents are unaware of how the DM will aggregate their recommendations. See Subsection~\ref{sub:str} for further discussion of this issue.} The DM receives the statistics of the survey -- i.e., the number of agents recommending implementation. A (probabilistic) \emph{aggregation rule} maps the survey's outcome to the probability with which the project will be implemented. Importantly, the aggregation rule can rely on knowledge of the marginal distributions of posteriors, but not on the correlation among the agents.

Our goal is to optimize the aggregation rule under the robust (adversarial) approach. On the one hand, the \emph{minimax} approach aims to find an aggregation rule that performs well for \emph{all possible correlations between the private information of different agents}. The minimax-optimal aggregation rule has been studied by~\citet{de2021robust}, who show that the optimal aggregation rule is to follow the best expert. On the other hand, the \emph{regret minimization} approach considers as a benchmark a hypothetical \emph{Bayesian DM}, who knows the true correlations between the agents' private information; given the statistics of the survey, the Bayesian DM can calculate the posterior distribution of the state, and take an optimal action given the survey outcome. The goal is to minimize the difference between the utility of the Bayesian DM and that of our ignorant DM who does not know the correlations.

The assumption of symmetric agents is common in many contexts~\cite[see, e.g.,][]{BulowK96}; in our leading example, this assumption is reasonable since the agents are ordinary citizens who obtain exogenous information with an unknown reliability level. Another possible interpretation of our model involves a researcher (DM) repeatedly experimenting -- making the symmetry assumption intrinsic. The knowledge of the marginals corresponds to the knowledge of the false-positive and the false-negative rates. Robustness to the correlation may be viewed as robustness to the sampling procedure according to which the samples for the experiments are collected.
\subsection{Outline of the Main Results}
We begin by revisiting the work of~\citet{de2021robust} on minimax-optimal aggregation and developing an analogue of their main result for symmetric agents. According to their result, the aggregator should follow the best expert. In our symmetric version, the DM only needs to observe the aggregate statistics of the agents' recommendations. Even though the aggregator cannot follow any single agent (their recommendations are treated anonymously), the aggregator can follow a random agent by adopting a recommendation chosen uniformly at random (the random dictator rule). Since, by symmetry, all agents are qualified to be the best experts, such an aggregation rule is minimax-optimal.

Turning from minimax to regret, the analysis is more involved. One challenge for the aggregation task is that the number of possible posterior belief realizations grows exponentially in the number of agents. This is where treating the agents anonymously is helpful. We utilize a recent characterization by~\cite{arieli2022population} of feasible information structures to reduce the problem to a single-dimensional one. In contrast to some cases with a small number of agents (as we demonstrate for two agents in Example~\ref{ex:two}), we prove that for a large class of instances -- in particular, for any sufficiently large number of agents -- the unique regret-minimizing aggregation function is the random dictator rule (Theorem~\ref{thm:dictator}). To deal with the remaining instances, we apply a zero-sum game formulation to deduce a clean formula for the minimal regret the DM can guarantee (Theorem~\ref{thm:gap}). The optimal regret turns out to be the maximum distance between a high-dimensional function $P^*$ representing the expected utility of a Bayesian DM and its concavification. This result implies a connection between minimizing regret and the concavification technique that is usually applied in Bayesian persuasion~\cite{kamenica2011bayesian}. Interestingly, even though our DM is ignorant of the information structure, her regret is tightly related to the probability with which a Bayesian DM correctly guesses the state of nature. The concavification formula allows us to resolve the regret problem in instances in which the previous approach fails, i.e., cases with a small number of agents.

While the anonymity assumption (by which the DM observes only the aggregate statistics of the agents' recommendations) is natural in the context of information aggregation, we also show that considering this anonymous version of the problem is without loss of generality (Subsection~\ref{sub:anonymity}).
\subsection{Related Work}
\label{sub:related}
The most closely related paper to ours is~\cite{de2021robust}. The settings considered in the two works are similar. The major difference is that our main focus is on the regret approach, which differs (conceptually and technically) from their minimax approach. Another difference is that we consider symmetric agents. Unlike~\cite{de2021robust}, we also study an anonymous variant of the problem in which the aggregator observes the statistics of recommendations rather than the entire recommendation vector.

A work that adopts the regret approach is~\citet{arieli2018robust}. They study Blackwell-ordered~\cite{blackwell1950comparison, blackwell1953equivalent} and conditionally i.i.d.~information structures. Additional works that consider regret as a measure of robustness  include:~\citet{babichenko2021learning}, which explores the class of partial-evidence information structures in a repeated game setting;~\cite{chen2016informational}, which considers information structures with informational substitutes; and~\citet{neyman2022you}, which shows that under the projective substitutes assumption, the principal can improve upon the random dictator mechanism considered by us.

A recent paper on robustness by~\citet{bogomolnaia2023guarantees} studies the guarantees of agents in a setting with unanimously adversarial other agents. Each agent has ordinal preferences over all possible outcomes, which may differ between agents. However, the model treats the agents symmetrically and anonymously. Unlike our paper, the agents compete each against the others, and there is no exogenous adversary. They prove that for any number of agents, each agent's best guarantee is the uniform distribution over the outcomes. Nevertheless, when the number of agents is smaller than the number of outcomes, they show that several variations of the random dictator rule and voting by veto are also optimal. Unlike in their paper, in our setting random dictator is uniquely optimal for a large enough number of agents.

Information aggregation with uncertainty about the information structure has been studied in a voting framework by~\citet{levy2015correlation}. They consider a voter's uncertainty about whether signals from different sources are fully correlated or conditionally independent. They show that such uncertainty may encourage the voter to rely more on the information she gets and less on her political preference. Robustness to correlations for fixed marginal distributions over agents' private types has been studied in the framework of mechanism design. Some examples~\cite{carroll2017robustness,he2022correlation} study the optimal robust mechanism for a principal, assuming that she knows the marginal distributions of the agents' types, but not the joint~distribution.

Forecast aggregation is a well-studied topic within statistics. Some works (see, e.g.,~\cite{degroot1974reaching,stock2004combination,zarnowitz1984accuracy}) examine how well simple aggregation rules perform. An alternative approach uses a training set to find an optimal aggregation scheme within a parametric family of such schemes (e.g.,~\cite{newbold1974experience}). 
In social choice, \emph{judgment aggregation} studies the aggregation of beliefs about whether logical statements are true or false~\cite{Endriss16}.
In economics,~\citet{levy2021maximum} consider forecast aggregation with multiple experts. Their aggregation scheme follows the assumption that the information structure of the experts is the one maximizing the likelihood of observing their realized forecasts. Further recent works on forecast aggregation include~\citet{WangLC21}, who study aggregation without being exposed to past performance data using peer assessment scores. In the correlation-robust framework, we should mention~\cite{levy2020combining}, who study the aggregation of information with uncertainty both about the joint information distribution and the set of all possible signals.

The fundamental question of whether social choice rules aggregate well the information of the crowd is an old topic in economics that goes back to~\citet{condorcet1785essai}, who proves that the simple majority rule is an asymptotically optimal aggregation rule when agents vote sincerely according to their private information. However,~\citet{austen1996information} demonstrate that Condorcet theorem strongly relies on the property that agents' information is symmetric. In asymmetric information structures, even when the signals are conditionally independent, the simple majority aggregation rule is far from optimal. These observations have initiated an extensive line of research that aims to understand which aggregation rules aggregate well the information of the crowd (e.g.,~\cite{mclennan1998consequences,feddersen1998convicting,laslier2013incentive,ahn2016approval}, just to mention a few). Our paper aims to address the same question but in a setting with the correlation of signals being unknown. This is in contrast to most of the existing research, which commonly assumes that conditional on the state the signals of the agents are independent. Our result indicates the robust optimality of the random dictator rule, and we show in Section~\ref{sec:concav} that in our symmetric setting, simple majority aggregation is far from optimal.

\paragraph{Paper organization.}
In Section~\ref{sec:preliminaries}, we introduce the formal model and basic definitions. The model includes some simplifying assumptions; generalizations are discussed in Section~\ref{sec:extensions}. Section~\ref{sec:minimax} revisits the minimax robustness paradigm in our setting, showing that the random dictator rule is optimal. In Section~\ref{sec:concav}, we present an approach for analyzing regret minimization that is useful for a large number of agents. This approach allows to bound from above the regret of any specific aggregation rule; we extend the main result of this section to the approximation ratio robustness paradigm in Subsection~\ref{sub:approx}. Section~\ref{sec:equivalence} connects regret-minimizing aggregation to the concavification of a function; Subsection~\ref{sub:two} shows an application for two agents. Section~\ref{sec:conclusion} concludes and suggests some questions for future work.
\section{Preliminaries}
\label{sec:preliminaries}
\subsection{Setting} 
\label{sub:sec}
Consider a binary space of \emph{states of nature} $\Omega=\braces*{0,1}$, equipped with a publicly-known \emph{prior} $\mu\in \brackets*{0,1}$ -- the probability that the (unknown) true state $\omega\in\Omega$ is equal to $1$. We extend our main results to an arbitrary finite state space in Subsection~\ref{sub:states}.

The model contains a \emph{decision maker} (or \emph{DM}, also known as an \emph{aggregator}) and $n$ informed \emph{agents}.
The agents are numbered $1,\ldots,n$, with agent $i$ observing a binary \emph{signal} $s_i\in S:=\braces*{L,H}$ that represents information about the state of nature. The agents truthfully report to the DM either $H$ or $L$ according to their observed signal. The assumption that signals are binary is for simplicity of presentation -- our results can be extended to arbitrary sets of signals, including continuum-many signals (see Subsection~\ref{sub:posteriors}). For a set $D$, let $\Delta\parentheses*{D}$ be the set of all distributions over~$D$. The joint distribution according to which the state and the $n$ signals are generated -- the \emph{information structure} -- is denoted by $\pi\in \Delta\parentheses*{\Omega \times S^n}$. We use $\pi\parentheses*{\cdot}$ to denote the probabilities according to the information structure. The information structure must be compatible with the prior distribution -- i.e., $\pi\parentheses*{\omega=1}=\mu$. 

\paragraph{The posteriors.}
Upon observing $s_i=\sigma$, agent $i$ computes her \emph{posterior distribution} over the states as follows: the posterior probability of $\omega=1$ is $\frac{\pi\parentheses*{\omega=1,s_i=\sigma}}{\pi\parentheses*{s_i=\sigma}}$. The agents are \emph{symmetric} in the following sense -- the marginal distributions over posterior beliefs of all the agents are identical. The posterior of agent $i$ after observing the signal $s_i=L$ is assumed to be fixed across all agents. Denote this posterior by $p_1$ and call it the \emph{low posterior}. Similarly, the \emph{high posterior} of an agent is $p_2$. Note that for every $1\le i\le n$ we have: $p_1=\frac{\pi\parentheses*{\omega=1,s_i=L}}{\pi\parentheses*{s_i=L}}; ~ p_2=\frac{\pi\parentheses*{\omega=1,s_i=H}}{\pi\parentheses*{s_i=H}}$.

\noindent Throughout the paper, we assume $p_1<\frac{1}{2}<p_2$ -- otherwise, the agents' reports are fixed, and the setting is degenerate. As in~\cite{de2021robust}, we assume that $p_1,p_2$ are known to the DM; this enables us to focus on the effect of not knowing the correlations, which are captured by the information structure~$\pi$ that is not observed by the DM.

Throughout the paper, we use the following parameters $a,b$: Parameter $a$ (respectively, $b$) represents the probability of a single fixed agent to have the high posterior conditional on state $\omega=0$ (respectively, $\omega=1$). The parameters are deduced by straightforward calculations to be:
\begin{equation}
a=\frac{\parentheses*{1-p_2}\parentheses*{\mu-p_1}}{\parentheses*{1-\mu}\parentheses*{p_2-p_1}}; ~
b=\frac{p_2\parentheses*{\mu-p_1}}{\mu\parentheses*{p_2-p_1}}. \label{eq:a-and-b} 
\end{equation}
\subsection{Information Aggregation} 
We consider an identical-interest scenario. Therefore, we assume that the agents truthfully report their private signals to the DM. However, we assume that the DM only observes the fraction of agents who report $H$, which is denoted by $\nu\in \braces*{0,\frac{1}{n},\ldots,1}$. Following~\cite{arieli2022population}, we assume \emph{anonymous} agents -- the DM does not observe the identity of the agents who report $H$. Note that \emph{this assumption is without loss of generality} (see Subsection~\ref{sub:anonymity}). It is introduced both for simplicity of presentation and for emphasizing the unimportance of the agents' identities for understanding the optimal DM'~strategy.

The DM's goal is to aggregate the information into a guess of the correct state, using her knowledge of the fraction $\nu$, as well as the prior $\mu$ and posteriors $p_1,p_2$. Namely, the DM's utility is $1$ if she guesses the state correctly, and is $0$ otherwise.\footnote{For a binary state space, it is easy to generalize all our results to arbitrary utilities; we further generalize our main results to any finite state space and arbitrary utilities in Subsection~\ref{sub:states}.} The DM can use a mixed guessing strategy, captured by an \emph{aggregation rule} $f:\braces*{0,\frac{1}{n},\ldots,1}\to \brackets*{0,1}$, where $f\parentheses*{\nu}$ denotes the probability that the DM guesses $\omega=1$ after observing a fraction of $\nu$ agents who report $H$.

The information structure $\pi$ and state of nature $\omega$ induce a distribution over $\nu$ -- that is, a distribution over the possible fractions $\braces*{0,\frac{1}{n},\ldots,1}$ of agents with the high posterior. Let $\hat{\pi}\in \Delta\parentheses*{\Omega \times \braces*{0,\frac{1}{n},\ldots,1}}$ denote the distribution over $\omega$ and $\nu$ induced by $\pi$. The corresponding distributions over $\nu$ conditional on the state $\omega=0$ and $\omega=1$ are denoted by $\hat{\pi}_0$ and $\hat{\pi}_1$, respectively. For an information structure $\pi$ and an aggregation rule $f$, the probability of the DM to guess correctly is denoted by $P\parentheses*{f,\pi}$, and equals:
\begin{align*}
    P\parentheses*{f,\pi}:=\mu \mathbb{E}_{\nu \sim \hat{\pi}_1} \brackets*{f\parentheses*{\nu}} + \parentheses*{1-\mu}\mathbb{E}_{\nu \sim \hat{\pi}_0} \brackets*{1-f\parentheses*{\nu}}.
\end{align*}
Note that $\mathbb{E}_{\nu \sim \hat{\pi}_1} \brackets*{f\parentheses*{\nu}}$ is the probability that the DM guesses $\omega=1$, on average over the observed fraction $\nu$ given that the true state is $1$. Similarly, $\mathbb{E}_{\nu \sim \hat{\pi}_0} \brackets*{1-f\parentheses*{\nu}}$ is the probability that the DM guesses $\omega=0$, on average over the observed fraction $\nu$ given that the true state is $0$.
\subsection{Regret and Minimax}
\label{sub:regret}
Consider a \emph{Bayesian DM}, i.e., a DM who knows the information structure $\pi$. Such a DM always guesses the more likely state. Hence, her probability of guessing correctly~equals:
\begin{align*}
    P^*\parentheses*{\pi}:=\mathbb{E}_{\nu \sim \hat{\pi}} [\max \braces*{\hat{\pi}\parentheses*{\omega=0|\nu}, \hat{\pi}\parentheses*{\omega=1|\nu}}].
\end{align*}
The \emph{regret of an aggregation rule $f$ for an information structure $\pi$} is defined by:
$$
\reg\parentheses*{f,\pi}:=P^*\parentheses*{\pi}-P\parentheses*{f,\pi}.
$$

Since the DM does not know $\pi$, she aims to design an aggregation rule with low regret \emph{for all} information structures.
We define $\reg\parentheses*{f}:=\max_\pi \reg\parentheses*{f,\pi}$.\footnote{The maximum exists as $\reg\parentheses*{f,\cdot}$ is a continuous functional defined on a compact space.} That is, the \emph{regret of an aggregation rule $f$} is the regret -- i.e., expected additive utility loss -- in the worst-case scenario. We refer to the worst-case information structure $\pi$ as chosen by an \emph{adversary} given $f$.
Denote by $\reg:=\min_f \reg\parentheses*{f}$ the regret of an \emph{optimal} aggregation rule. Through most of the paper, we focus on computing $\reg$ and finding an aggregation rule $f$ (which may depend on $p_1,p_2$, and $\mu$, but not on $\pi$) that minimizes $\reg \parentheses*{f}$.

We further define 
$$
\minmax:=\max_{f}\min_{\pi} P\parentheses*{f,\pi}
$$ 
to be the maximum probability of a correct guess by the DM under the worst possible information structure chosen by an adversary. We say that an aggregation rule $f$ is \emph{minimax-optimal} if $\min_{\pi} P\parentheses*{f,\pi}=\minmax$.
\subsection{Concavification and Convexification}
\label{sec:con}
We recall the standard definitions of the concavification and convexification of a function~\cite{blackwell1953equivalent,aumann1995repeated}. Let $D$ be some compact convex set and let $h:D\to \mathbb{R}$ be a function. The \emph{concavification} $\cav\brackets*{h}:D\to \mathbb{R}$ is the pointwise minimum of all the concave functions whose graph is weakly above $h$. Formally, $\cav\brackets*{h}$ is concave and for every concave function $\tilde{h}:D\to \mathbb{R}$ with $\tilde{h}\parentheses*{x}\geq h\parentheses*{x}$ for every $x\in D$, it holds that $\tilde{h}\parentheses*{x}\geq\cav\brackets*{h}\parentheses*{x}\geq h\parentheses*{x}$ for every $x\in D$.\footnote{See~\cite{dughmi2009submodular} for algorithmic aspects of computing the concavification of a set function.} Similarly, the \emph{convexification} $\vex\brackets*{h}:D\to \mathbb{R}$ is the pointwise maximum of all convex functions weakly below~$h$.
\section{Minimax-Optimal Aggregation Revisited}
\label{sec:minimax}
In this section, we revisit the work of~\citet{de2021robust}, who study minimax-optimal information aggregation rules. Unlike us, they do not assume that the agents' distributions over posteriors are identical. The main result of~\cite{de2021robust} states that the minimax-optimal aggregation rule is to follow the recommendation of the best agent (and ignore the remaining agents); ``best agent'' here means the agent with the most informative private signal. We now give an analogue of this result in our symmetric setting and provide a simple standalone proof for completeness.

The simple key observation in our symmetric setting is as follows. Since all agents have identical distributions of posteriors, they all qualify as ``best agents''. By choosing an agent uniformly at random and following her recommendation (an aggregation rule for which anonymous statistics suffice), we essentially follow the recommendation of the best agent. \footnote{We assume the DM only observes anonymous data. As we show in Subsection~\ref{sub:anonymity} for regret minimization -- this assumption is not essential. Analogous considerations apply to minimax-optimal aggregation rules.}
Therefore, the minimax-optimal aggregation rule is the \emph{random dictator} rule -- choosing an agent uniformly at random and relying only on her recommendation.~Formally:
\begin{definition}
    \label{def:dictator}
    The \emph{random dictator} aggregation rule is the function $f:\braces*{0,\frac{1}{n},\ldots,1}\to\brackets*{0,1}$ satisfying $f\parentheses*{\frac{k}{n}}=\frac{k}{n}$ for $0\leq k\leq n$.
\end{definition}
\noindent For example, if the fraction of agents with signal $H$ is $\nu=1/3$, the probability of guessing $\omega=1$ is $f\parentheses*{1/3}=1/3$, which is equivalent to choosing one of the $n$ agents uniformly at random and following her recommendation.

In the symmetric setting we study, one can analyze optimal aggregation for the minimax paradigm straightforwardly, without relying on the results of~\cite{de2021robust}. It is informative to consider in this case the best information structure from the perspective of an adversary who tries to fail the DM. Such an information structure is \emph{fully-correlated}, that is, it sends all the agents the same signal, thus revealing the minimal possible amount of information to the DM. The DM then has nothing better to do than follow the unanimous recommendation of all agents.
\begin{proposition}[A special case of the main result of~\cite{de2021robust}]
    \label{pro:minimax}
    For every information aggregation setting with prior $\mu$, the random dictator aggregation rule is minimax-optimal. Moreover, $\minmax=\parentheses*{1-\mu}\cdot\parentheses*{1 - a}+ \mu\cdot b$ for $a,b$ as defined in Eq.~\eqref{eq:a-and-b}.
\end{proposition}
\begin{proof}
    We first prove that the random dictator rule guesses $\omega$ correctly with expected probability of exactly $\parentheses*{1-\mu}\cdot\parentheses*{1 - a} + \mu\cdot b$. Indeed, when $\omega=0$, the expected fraction of agents with the low posterior is $1-a$. Moreover, when $\omega=1$, the expected fraction of agents with a high posterior is $b$. Since the DM follows the recommendation of a uniformly chosen agent, the law of total expectation yields that the expected probability of correct guess is $\parentheses*{1-\mu}\cdot\parentheses*{1 - a} + \mu\cdot b$, as desired.

    It remains to show that for every aggregation rule $f$, the adversary can ensure that the probability of a correct guess is at most $\parentheses*{1-\mu}\cdot\parentheses*{1 - a} + \mu\cdot b$. Indeed, let the adversary pick the fully-correlated information structure -- that is, $s_1\equiv s_2\equiv\ldots\equiv s_n$. In this case, the random dictator rule is trivially optimal, as all the agents get the same signal. Since, as shown, this rule guesses the correct state with an expected probability of exactly $\parentheses*{1-\mu}\cdot\parentheses*{1 - a} + \mu\cdot b$, the claim~follows.
\end{proof}
This straightforward analysis no longer holds for the regret minimization paradigm. Indeed, for the fully-correlated information structure, our ignorant DM can perform as well as the hypothetical Bayesian DM, and the regret is $0$. The next two sections provide two complementary approaches to tackling regret minimization.
\section{Main Result}
\label{sec:concav}
In this section, we present a general approach for finding the minimum regret $\reg$, which -- quite surprisingly -- turns the optimization into a \emph{single-dimensional} problem. We prove that for many settings, the uniquely optimal aggregation rule $f$ is the random dictator rule (Theorem~\ref{thm:dictator}). In particular, it is always true for a sufficiently large number of agents $n$ (Corollary~\ref{cor:asymp}). We start with a preliminary bound on the regret of a given aggregation rule $f$; see Figure~\ref{fig:conv-conc}.
\begin{figure}[h]
    \centering
    \scalebox{0.7}{%
    \begin{tikzpicture}
    \draw[->] (-0.1,0)--(10.1,0);
    \draw[->] (0,-0.1)--(0,5.1);
    \node[right] at (10.1,0) {$\nu$};
    \node[above] at (0,5.1) {$f$};
    \foreach \x in {1,...,10}{
    \draw (\x,-0.1)--(\x,0.1);
    }
    \node[below] at (1,-0.1) {$\frac{1}{n}$};
    \node[below] at (2,-0.1) {$\frac{2}{n}$};
    \node[below] at (9,-0.1) {$\frac{n-1}{n}$};
    \node[below] at (10,-0.1) {$1$};
    
    \draw[dashed] (0,5)--(10,5);
    \node[left] at (0,5) {1};
    
    \filldraw (0,1.5) circle(0.07);
    \filldraw (1,0.5) circle(0.07);
    \filldraw (2,3) circle(0.07);
    \filldraw (3,2.5) circle(0.07);
    \filldraw (4,1) circle(0.07);
    \filldraw (5,3.5) circle(0.07);
    \filldraw (6,4) circle(0.07);
    \filldraw (7,3) circle(0.07);
    \filldraw (8,3.5) circle(0.07);
    \filldraw (9,3.6) circle(0.07);
    \filldraw (10,4.5) circle(0.07);
    
    \draw[blue] (0,1.5)--(2,3)--(6,4)--(10,4.5);
    \draw[red] (0,1.5)--(1,0.5)--(4,1)--(9,3.6)--(10,4.5);
    
    \node[above] at (2,3) {$\underline{a}$};
    \node[above] at (6,4) {$\overline{a}$};
    
    \node[below] at (4,1) {$\underline{b}$};
    \node[below] at (9,3.6) {$\overline{b}$};
    
    \draw (3.5,-0.1)--(3.5,0.1);
    \draw (6.5,-0.1)--(6.5,0.1);
    \node[below] at (3.5,-0.1) {$a$};
    \node[below] at (6.5,-0.1) {$b$};
    
    \draw[line width=1.5, green!50!black] (3.5,3.38)--(3.5,5);
    \draw[line width=1.5, green!50!black] (6.5,2.3)--(6.5,0);
    
    \node[green!50!black,left] at (3.5,4.1) {$\omega=0$};
    \node[green!50!black,right] at (6.5,1.1) {$\omega=1$};
    
    \end{tikzpicture}}
    \caption{An aggregation rule $f$. The blue function is $\cav\brackets*{f}$. The red function is $\vex\brackets*{f}$. The length of the left (resp., right) green line captures the minimal probability -- across all information structures -- of the aggregation function $f$ to guess correctly the state $\omega=0$ (resp., $\omega=1$).}
    \label{fig:conv-conc}
\end{figure}
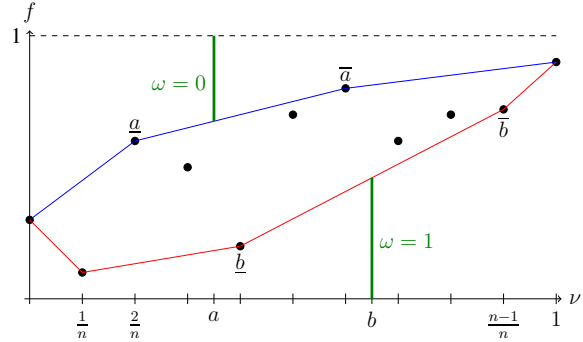
\begin{proposition}
\label{pro:concav}
For every $a,b,n,\mu,$ and $f$:
\begin{align*}
    &\reg \parentheses*{f}\leq 1-\parentheses*{1-\mu}\cdot\parentheses*{1- \cav\brackets*{f}\parentheses*{a}}-\mu\cdot \vex\brackets*{f}\parentheses*{b}.
\end{align*}
\end{proposition}
To prove this result, we first note that the set of feasible distributions over frequencies of posteriors has a neat characterization according to a recent result by~\citet{arieli2022population}. More formally, we say that a pair of conditional distributions $\parentheses*{\hat{\rho}_0,\hat{\rho}_1}\in \parentheses*{\Delta\parentheses*{\braces*{0,\frac{1}{n},\ldots,1}}}^2$ is \emph{feasible} if there exists an information structure $\pi$ such that $\hat{\rho}_\omega=\hat{\pi}_\omega$ for $\omega=0,1$. The parameters $a$ and $b$ defined in Section~\ref{sec:preliminaries} play a central role in the characterization of feasible~distributions.
\begin{theorem}[\citet{arieli2022population}]\label{thm:ab}
A pair of conditional distributions $\parentheses*{\hat{\rho}_0,\hat{\rho}_1}\in\parentheses*{\Delta\parentheses*{\braces*{0,\frac{1}{n},\ldots,1}}}^2$ is feasible if and only if $\mathbb{E}\brackets*{\hat{\rho}_0}=a$ and $\mathbb{E}\brackets*{\hat{\rho}_1}=b$.
\end{theorem}
\begin{proof}[Proof of Proposition~\ref{pro:concav}]
Let $\pi$ be an information structure that induces the distributions of frequencies $\parentheses*{\hat{\pi}_0,\hat{\pi}_1}$. By Theorem~\ref{thm:ab}, we know that $\mathbb{E}\brackets*{\hat{\pi}_0}=a$ and $\mathbb{E}\brackets*{\hat{\pi}_1}=b$. Since $\mathbb{E}\brackets*{\hat{\pi}_0}=a$, the function $f$ guesses the state with probability of at least $1-\cav\brackets*{f}\parentheses*{a}$. Indeed, $\cav\brackets*{f}$ captures the maximal probability of a mistake (across all possible $\hat{\pi}_0$). Similarly, as $\mathbb{E}\brackets*{\hat{\pi}_1}=b$, the function $f$ guesses the state with probability of at least $-\cav\brackets*{-f}\parentheses*{b}=\vex\brackets*{f}\parentheses*{b}$: $\vex\brackets*{f}$ captures the minimal probability of a correct guess. Therefore, the probability of a correct guess of $\omega$ under the information structure $\pi$ is at~least:
\begin{align*}
   1-\parentheses*{1-\mu}\cdot\parentheses*{1- \cav\brackets*{f}\parentheses*{a}}-\mu\cdot \vex\brackets*{f}\parentheses*{b}.
\end{align*}
With the trivial bound of 1 on the correct guess probability of the state, we get for every $\pi$:
\begin{align*}
    \reg \parentheses*{f,\pi}\leq 1-\parentheses*{1-\mu}\cdot\parentheses*{1- \cav\brackets*{f}\parentheses*{a}}-\mu\cdot \vex\brackets*{f}\parentheses*{b}.
\end{align*}
Since it is true for every $\pi$, the proposition follows.
\end{proof}

\paragraph{Tightness of Proposition~\ref{pro:concav}.} The bound of Proposition \ref{pro:concav} is tight in many cases. 
For instance, consider the function $f$ that is depicted in Figure~\ref{fig:conv-conc}. By Theorem \ref{thm:ab}, the adversary can choose the information structure $\pi$ such that $\hat{\pi}_0$ will be supported on the two points $\underline{a},\overline{a}$, and $\hat{\pi}_1$ will be supported on the two points $\underline{b},\overline{b}$. For such an information structure, the Bayesian DM knows to guess the state with probability 1 because $\braces*{\underline{a},\overline{a}}\cap \braces*{\underline{b},\overline{b}}=\emptyset$. Moreover the aggregation function $f$ guesses the state $\omega=0$ (respectively, $\omega=1$) with probability $1-\cav\brackets*{f}\parentheses*{a}$ ($\vex\brackets*{f}\parentheses*{b}$) exactly. Hence, the bound of Proposition~\ref{pro:concav} is tight. In fact, such an argument will prove the tightness of the bound in Proposition~\ref{pro:concav} whenever the supports of the concavification of $f$ at $a$ and of the convexification of $f$ at $b$ are \emph{disjoint}. This is a key observation to deduce the regret-minimizing $f$ for a large class of instances in the case of many agents (i.e. when $n$ is large).

\paragraph{Regret minimization of the random dictator rule.} We prove that if $\frac{1}{n}\leq a < b \leq \frac{n-1}{n}$ -- then the uniquely optimal aggregation rule is random dictator. As we have already noted in Section~\ref{sec:minimax}, this aggregation rule is the natural analogue in our symmetric setting of ``following the best agent", which is shown to be minimax-optimal by~\cite{de2021robust}. Moreover, it follows that 
random dictator is the \emph{unique approximation ratio-optimizing aggregation rule} when $\frac{1}{n}\leq a < b \leq \frac{n-1}{n}$. Thus, for fixed prior and marginal posteriors, the random dictator rule is asymptotically uniquely optimal \emph{regardless of the studied robustness paradigm}.
\begin{theorem}[Main]
\label{thm:dictator}
Assume $\frac{1}{n}\leq a < b \leq \frac{n-1}{n}$. Then random dictator is the unique regret-minimizing aggregation rule, and~$\reg=1-\parentheses*{1-\mu}\parentheses*{1-a}-\mu b$.
\end{theorem}
In particular, this theorem implies an asymptotic result when the marginals remain fixed, but the population of agents grows.
\begin{corollary}
\label{cor:asymp}
For every $\mu, p_1$ and $p_2$, there exists $N$ s.t. for every $n\geq N$, the random dictator rule is the unique regret-minimizing aggregation rule. Specifically, $N=\max\braces*{\frac{\parentheses*{1-\mu}\parentheses*{p_2-p_1}}{\parentheses*{1-p_2}\parentheses*{\mu-p_1}},\frac{\mu\parentheses*{p_2-p_1}}{p_1\parentheses*{p_2-\mu}}}$.
\end{corollary}
In particular, $N$ is large when one of the marginal posteriors is close to an extreme point of the interval to which it might belong -- the prior, $0$ or $1$. Intuitively, when one of the possible signals either reveals almost no information or reveals almost full information -- some intricate aggregation rules might be required, giving, respectively, less or more weight to the recommendations of the agents with this extreme signal. Intuitively, the proof of the theorem relies on the tightness of Proposition~\ref{pro:concav} discussed above. However, showing that there exists a best signaling scheme $\pi^*$ for the adversary s.t.~the supports of $\hat{\pi^*}_0,\hat{\pi^*}_1$ are disjoint turns out to be tricky. We first show that whenever the constants $a$ and $b$ lie in the segment $\brackets*{\frac{1}{n},\frac{n-1}{n}}$ -- there exists an optimal aggregation rule that is a \emph{linear} function. This is because the adversary has sufficient flexibility to choose an information structure $\pi$ with disjoint supports for $\hat{\pi}_0$ and $\hat{\pi_1}$ that will drop the DM's probability of guessing (weakly) below the probability achieved by some linear function. Because of the disjointness property, such an information structure has an ideal performance for the Bayesian DM (i.e., it always guesses the state correctly). Secondly, we prove that the optimal linear aggregation function is the random dictator. The full proof of Theorem~\ref{thm:dictator} appears in Appendix~\ref{app:dictator}.

\paragraph{Supermajority aggregation.} A common class of aggregation functions that are considered both in practice and in the theoretical literature is the class of \emph{supermajority rules}. Namely, $f\parentheses*{\nu}=\mathbf{1}_{\nu\geq \tau}$ for some threshold $\tau$. We note that supermajority rules might perform very badly in terms of regret. For example, fix $a=\frac{1}{2}-\frac{1}{n},b=\frac{1}{2}+\frac{1}{n}$ and $\mu=\tau=\frac{1}{2}$. The regret of the majority rule is as high as $1-O\parentheses*{\frac{1}{n}}$; namely, there exists an information structure for which the Bayesian aggregator guesses the state with probability $1$, while the majority rule guesses the state with probability $O\parentheses*{\frac{1}{n}}$. To show this, one may consider an information structure for the adversary with $\hat{\pi}_0$ supported on $0$ and a grid point $\nu_0\in \parentheses*{\frac{1}{2},\frac{1}{2}+O\parentheses*{\frac{1}{n}}}$, and with $\hat{\pi}_1$ supported on $1$ and a grid point $\nu_1\in \parentheses*{\frac{1}{2}-O\parentheses*{\frac{1}{n}},\frac{1}{2}}$. Note that conditional on state $\omega$, with probability $1-O\parentheses*{\frac{1}{n}}$ the realized fraction of high reports will be $\nu_\omega$, which will cause the majority rule to fail. In contrast, the random dictator rule will guess the state correctly with a probability of $\frac{1}{2}+\frac{1}{n}$.

\paragraph{Alternative robustness approaches.}
Proposition~\ref{pro:concav} proof immediately implies a stronger result than Proposition~\ref{pro:minimax} -- under the minimax paradigm, the random dictator is always the \emph{unique} optimal aggregation rule. Moreover, it follows from Theorem~\ref{thm:dictator} proof that whenever $\frac{1}{n}\leq a<b\leq\frac{n-1}{n}$, the adversary has an optimal strategy that allows the Bayesian DM to guess the state correctly with probability $1$. Therefore, the regret-minimizing aggregation rule is also approximation ratio-optimizing -- see Subsection~\ref{sub:approx}. Hence, given any fixed prior and marginal posteriors, for a large enough number of agents $n$, \emph{the random dictator rule is uniquely optimized for all three robustness paradigms: minimax, regret and approximation ratio}. It highlights the universal nature of the robust optimality of random~dictator.
\section{General Regret Analysis}
\label{sec:equivalence}
In the previous section, we showed how to compute the minimal regret for a large number of agents. In this section, we characterize the minimal regret as the maximal difference between the concavification of a function $P^*$ -- representing the correct guess probability of the Bayesian DM -- and $P^*$ itself (Theorem~\ref{thm:gap}). The formula uses the concavification of the function $P^*$ defined on a $2^{n+1}$-dimensional set; it allows explicitly to compute $\reg$ for a small number of agents.

Define, as is standard, $\brackets*{n}:=\braces*{1,\ldots,n}$. Note that the set of information structures with the given prior $\mu$ and given posteriors $p_1,p_2$ for both agents is a polygon $C\subset \mathbb{R}^{2^{n+1}}$ that is defined by the following equations:
\begin{align*}
    C:=\{x=\parentheses*{x^\omega_D}_{D\subseteq \brackets*{n},\;\omega=0,1} : & x^\omega_D \geq 0, \sum_{D\subseteq \brackets*{n}} x_{D}^0=1-\mu, \sum_{D \subseteq \brackets*{n}} x_{D}^1=\mu,\\
    & \forall i\in \brackets*{n}: \sum_{D:\;i\in D} x^0_D=\parentheses*{1-\mu}a, \sum_{D:\;i\in D} x^1_D=\mu b\},
\end{align*}
where the terms $\parentheses*{1-\mu}a$ and $\mu b$ represent the \emph{unconditional} probability weight that is assigned to the high signal at states $0$ and $1$, correspondingly. We now present a closed formula for the optimal~regret. We slightly abuse notation and consider $P^{*}\parentheses*{\pi}$, the probability of a correct guess by a Bayesian DM, to be a function $P^{*}: C\to \brackets*{0,1}$. Note that we have for every $x=\parentheses*{x^\omega_D}_{D\subseteq \brackets*{n},\;\omega=0,1}\in C$:
\begin{align*}
    P^{*}\parentheses*{x}=\sum_{i=0}^n \max \braces*{ \sum_{D:\;\absolute*{D}=i} x^0_D, \sum_{D:\;\absolute*{D}=i} x^1_D}.
\end{align*}
Note that $P^{*}$ is a convex function, as a sum of maxima of linear functions. The following theorem characterizes optimal regret. It is somewhat surprising as it connects the regret of a DM \emph{ignorant} of the information structure to the correct guess probability by a Bayesian DM \emph{aware} of the information~structure. 
\begin{theorem}
\label{thm:gap}
For every number of agents $n\geq 2$ it holds that $\reg=\max_{x\in C}\brackets*{\cav\brackets*{P^{*}}\parentheses*{x}-P^{*}\parentheses*{x}}$.\footnote{Theorem~\ref{thm:gap} proof does not rely on the agents' symmetry -- the theorem continues to hold for agent-dependent sets of posteriors $\braces*{p_1^i, p_2^i}$ ($1\leq i\leq n$). The definitions of $C$ and the function $P^*$ are similar to the symmetric case. An analogue of this theorem is valid in a more general setting without the symmetry assumption on the agents -- the setting considered by~\cite{de2021robust}; every agent has a different information structure and reports a recommendation. The probability of a correct guess by a Bayesian DM  $P^*$ is well-defined, and the theorem states that $\reg=\max_{x\in C}\brackets*{\cav\brackets*{P^{*}}\parentheses*{x}-P^{*}\parentheses*{x}}$. The validity of this generalization follows from the theorem proof relying on max-min arguments only, which hold in the non-symmetric case without modifications.}
\end{theorem}
\begin{proof}
The regret minimization problem is a zero-sum game between the DM who chooses an aggregation rule $f$ and an adversary who chooses the information structure $\pi$ (or, equivalently, $x\in C$). Concretely, deterministic aggregation rules $f:\braces*{0,\frac{1}{n},\ldots,1}\to \braces*{0,1}$ are the pure strategies of the DM, and information structures $x\in C$ are the pure strategies of the adversary. The payoff as a function of the DM's mixed strategy $f$ and the adversary's mixed strategy $\phi$ is:
\begin{align*}
     &\reg\parentheses*{f,\phi}:=\mathbb{E}_{x\sim\phi}\Bigg[\sum_{i=0}^n\left(\max\braces*{ \sum_{D\subseteq\Omega:\;\absolute*{D}=i} x^0_D, \sum_{D\subseteq\Omega:\;\absolute*{D}=i} x^1_D}\right. \nonumber\\
     &\qquad \left.-\parentheses*{\parentheses*{1-f_i}\cdot \sum_{D\subseteq\Omega:\;\absolute*{D}=i} X_D^0 +f_i \cdot \sum_{D\subseteq\Omega:\;\absolute*{D}=i} x_D^1}\right)\Bigg].
\end{align*}
The game value is $\reg$. By the minimax theorem, it equals the maximum over all mixed strategies of the adversary followed by a best-response of the DM. The DM's best response to a mixed strategy over information structures is her Bayesian guessing of a state given the \emph{known} mixed strategy of the adversary. Formally, let $\phi\in \Delta\parentheses*{C}$ be an adversary's mixed strategy. The DM's utility under best response is $P^{*}\parentheses*{\mathbb{E}\brackets*{\phi}}$. The Bayesian DM applies Bayesian guessing in each realization of $x\sim \phi$. Hence, her expected probability of guessing is $\mathbb{E}_{x\sim \phi} \brackets*{P^{*}\parentheses*{x}}$. Thus:
\begin{align*}
    \reg = \max_{\phi\in \Delta\parentheses*{C}} \brackets*{\mathbb{E}_{x\sim \phi} \brackets*{P^{*}\parentheses*{x}} - P^{*}\parentheses*{\mathbb{E}\brackets*{\phi}}} = \max_{y\in C} \brackets*{\cav\brackets*{P^{*}}\parentheses*{y}-P^{*}\parentheses*{y}},
\end{align*}
where the second equality is by taking $y=\mathbb{E}\brackets*{\phi}$ at the maximizing mixed strategy.
\end{proof}
\subsection{Example: The Two-Agent Case}
\label{sub:two}
In this subsection, we demonstrate the use of the duality method on which Theorem~\ref{thm:gap} is based for computing the best aggregation function for $n=2$ agents and a uniform prior. The result easily generalizes to an arbitrary prior, and a similar methodology can be used to find a closed expression for $\reg$ for small values of $n$.\footnote{Note that in the special case of the uniform prior, the assumption $p_1<\frac{1}{2}<p_2$ trivially holds unless~$p_1 = p_2 = a = b = \frac{1}{2}$.}
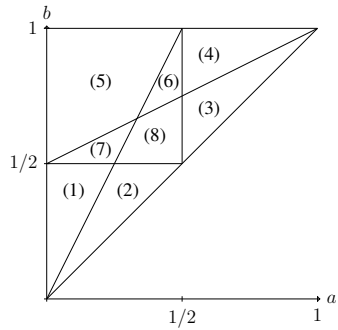
\begin{figure}[h]
    \centering
    \scalebox{0.6}{%
    \begin{tikzpicture}[scale=0.6]
    \draw[->] (-0.1,0)--(10.1,0);
    \draw[->] (0,-0.1)--(0,10.1);
    \node[right] at (10.1,0) {$a$};
    \node[above] at (0,10.1) {$b$};
    \draw (0,0)--(10,10);
    \draw (0,0)--(5,10);
    \draw (0,10)--(10,10);
    \draw (0,5)--(5,5);
    \draw (5,5)--(5,10);
    \draw (0,5)--(10,10);
    \draw (-0.1,5)--(0.1,5);
    \draw (-0.1,10)--(0.1,10);
    \draw (5,-0.1)--(5,0.1);
    \draw (10,-0.1)--(10,0.1);
    \node[below] at (5,-0.1) {$1/2$};
    \node[below] at (10,-0.1) {$1$};
    \node[left] at (-0.1,5) {$1/2$};
    \node[left] at (-0.1,10) {$1$};
    \node at (1,4) {(1)};
    \node at (3,4) {(2)};
    \node at (2,8) {(5)};
    \node at (2,5.5) {(7)};
    \node at (4,6) {(8)};
    \node at (4.5,8) {(6)};
    \node at (6,9) {(4)};
    \node at (6,7) {(3)};
    \end{tikzpicture}}
    \caption{The partition of $\parentheses*{a,b}$ into regions according to the cases in Example~\ref{ex:two}.}
    \label{fig:2p}
\end{figure}
\begin{example}
\label{ex:two}
For $n=2$ agents and a uniform prior $\mu=1/2$, the optimal aggregation function $f$ satisfies $f\parentheses*{0}=0$ and $f\parentheses*{1}=1$. The value of $f\parentheses*{\frac{1}{2}}$ and the minimal regret $\reg$ are given by:\footnote{When multiple cases apply -- all the corresponding possibilities for $f$ are optimal.}
\begin{enumerate}
 \item $f\parentheses*{\frac{1}{2}}=\frac{a+2b}{2\parentheses*{a+b}},\; \reg=\frac{a\parentheses*{a+2b}}{2\parentheses*{a+b}}$ if $a<b\leq \frac{1}{2}$ and $2a\leq b$.
 \item $f\parentheses*{\frac{1}{2}}=\frac{3b-a}{2\parentheses*{a+b}},\; \reg=\frac{a^2+4ab-b^2}{2\parentheses*{a+b}}$ if $a<b\leq \frac{1}{2}$ and $2a\geq b$.
 \item $f\parentheses*{\frac{1}{2}}=\frac{2-3a+b}{2\parentheses*{2-a-b}},\; \reg=\frac{-a^2+4ab+b^2-2a-6b+4}{2\parentheses*{2-a-b}}$ if $\frac{1}{2}\leq a<b$ and $1+a\geq 2b$.
 \item $f\parentheses*{\frac{1}{2}}=\frac{3-2a-b}{2\parentheses*{2-a-b}},\; \reg=\frac{\parentheses*{1-b}\parentheses*{3-2a-b}}{2\parentheses*{2-a-b}}$ if $\frac{1}{2}\leq a<b$ and $1+a\leq 2b$.
 \item $f\parentheses*{\frac{1}{2}}=\frac{1-b}{1+a-b},\; \reg=\frac{a\parentheses*{1-b}}{1+a-b}$ if $a\leq \frac{1}{2}\leq b$ and $b\geq \max\braces*{2a,\frac{a+1}{2}}$.
 \item $f\parentheses*{\frac{1}{2}}=\frac{2-2a-b}{2\parentheses*{1+a-b}},\; \reg=\frac{\parentheses*{1-b}\parentheses*{4a-b}}{2\parentheses*{1+a-b}}$ if $a\leq \frac{1}{2}\leq b$ and $2a \geq b\geq\frac{a+1}{2}$.
 \item $f\parentheses*{\frac{1}{2}}=\frac{-1+a+2b}{2\parentheses*{1+a-b}},\; \reg=\frac{a\parentheses*{3+a-4b}}{2\parentheses*{1+a-b}}$ if $a\leq \frac{1}{2}\leq b$ and $\frac{a+1}{2} \geq b\geq 2a$.
 \item $f\parentheses*{\frac{1}{2}}=\frac{3-a-3b}{2\parentheses*{1+a-b}},\; \reg=\frac{a^2-6ab+b^2+5a-b}{2\parentheses*{1+a-b}}$ if $a\leq \frac{1}{2}\leq b$ and $b\leq \min\braces*{2a,\frac{a+1}{2}}$.
\end{enumerate}
\end{example}

Example~\ref{ex:two} demonstrates that for a small number of agents ($n=2$), the regret-minimizing aggregation rule is quite a complex object (see Figure~\ref{fig:2p} for illustration). This is in sharp contrast to considering the minimax-optimal aggregation rule, which is always random dictator (Proposition~\ref{pro:minimax}). Note that in particular, there might be more than one regret-minimizing $f$. As follows from the previous section, this is a side effect of the number of agents being small.

The proof relies on Theorem~\ref{thm:gap}, which allows to calculate $\reg$ by maximizing the difference $\cav\brackets*{P^{*}}-P^{*}$. The function $P^*$ turns out to be two-dimensional; hence, the calculations are relatively easy. Note that Theorem~\ref{thm:gap} does not characterize the regret-minimizing aggregation rule $f$. To deduce $f$, we use the fact that $f$ must induce indifference between the information structures that the adversary chooses with positive probability. This pins down $f$ combined with the educated guess that $f\parentheses*{0}=0$ and $f\parentheses*{1}=1$. These calculations appear in Appendix~\ref{app:two}.

The analysis of Example~\ref{ex:two} in Appendix~\ref{app:two} suggests the following interpretation of the optimal $f$ for $n=2$ and $\mu=\frac{1}{2}$ -- the adversary should randomize between two information structures; in each of the two information structures, in one state the adversary should fully correlate the signals, while in the other state, she should anticorrelate the signals as much as~possible.
\section{Extensions}
\label{sec:extensions}
In this section, we discuss the generality of our results from previous sections. We start by showing that all our results hold even without the assumption of binary marginal posteriors (Subsection~\ref{sub:posteriors}). We proceed by showing that our main results on the optimality of random dictator (Theorem~\ref{thm:dictator}
and Corollary~\ref{cor:asymp}) hold for an arbitrary state space (Subsection~\ref{sub:states}). Then we establish that throughout the paper, one can drop the assumption that agents' reports to the DM are anonymous (Subsection~\ref{sub:anonymity}). Finally, we show how the proof of Theorem~\ref{thm:dictator} implies that for a sufficiently large number of agents, the random dictator aggregation rule also optimizes the approximation ratio (Subsection~\ref{sub:approx}). We conclude that for a large enough number of agents, the random dictator rule is the unique optimal aggregation rule regardless of the robustness paradigm -- minimax, regret, or approximation ratio. Finally, we discuss the truthfulness assumption on the agents in Subsection~\ref{sub:str}.
\subsection{Non-Binary Signals}
\label{sub:posteriors}
As we mentioned in Section~\ref{sec:preliminaries}, we assume that $\absolute*{S}=2$ purely for simplicity. Here, we show that all our results continue to hold in a much more general setting, provided that the agents still submit binary recommendations. We consider the model described in Section~\ref{sec:preliminaries} with the following differences. We take an arbitrary set of signals $S\subseteq \brackets*{0,1}$, with each signal labeled by the marginal posterior it induces on an agent observing it. We consider identical distributions over marginal posteriors $\pi_1,\ldots,\pi_n$ of agents no.~$1,\ldots,n$ (respectively), with $\pi_i\in\Delta\parentheses*{\Omega\times S}$ being the projection of $\pi\in\Delta\parentheses*{\Omega\times S^n}$ to the first and the $i+1$-th coordinates. We assume that $\min\braces*{\Pr_{q\sim\pi_1}\brackets*{q<1/2},\Pr_{q\sim\pi_1}\brackets*{q>1/2}}>0$, where $q$ denotes the marginal posterior probability for $\omega=1$.\footnote{If, e.g., all the elements of $S$ induce posteriors not below $1/2$ -- the DM clearly should just always guess~$\omega=1$.}

Each agent makes a report in $\braces*{H,L}$. Specifically, if her marginal posterior $q\in S$ satisfies $q\geq 1/2$ -- she chooses the \emph{high report ($H$)}, which is interpreted as $\omega=1$ being the likely possibility; otherwise, she chooses the \emph{low report ($L$)}. Similarly to the basic setting, our ignorant DM only observes the fraction of agents with the $H$ report, but not the report of each agent separately. Moreover, the DM knows the distribution over marginal posteriors, but not the full information structure $\pi$. In contrast, the \emph{Bayesian} DM observes the report of each agent separately, and she also knows the full information structure $\pi$.\footnote{We stress that neither the ignorant nor the Bayesian DM observe the exact marginal posteriors of the agents, but only their recommendations.} Denote $p_1:=\mathbb{E}_{q\sim \pi_1}\brackets*{q | q< 1/2}$ and $p_2:=\mathbb{E}_{q\sim \pi_1}\brackets*{q | q\geq 1/2}$ (both are well-defined by our assumption on the marginal distributions). Then $p_1<1/2<p_2$. We claim that all the proofs we presented in the paper are valid in this more general setting.

First, note that given an information structure $\pi$ in the general setting and the induced information structure $\pi'$ with the only possible marginal posteriors being $p_1, p_2$ -- each aggregation function $f$ satisfies:
\begin{align*}
    &\mathbb{E}_{\nu \sim \hat{\pi}_1} \brackets*{f\parentheses*{\nu}}=\mathbb{E}_{\nu \sim \hat{\pi'}_1} \brackets*{f\parentheses*{\nu}},\;\;\mathbb{E}_{\nu \sim \hat{\pi}_0} \brackets*{1-f\parentheses*{\nu}}=\mathbb{E}_{\nu \sim \hat{\pi'}_0} \brackets*{1-f\parentheses*{\nu}}.
\end{align*}
Indeed, the ignorant DM can emulate any aggregation function from the basic setting in the new setting too without making any changes, and vice versa. Denote by $s^B\parentheses*{\cdot}$ the state that the Bayesian DM guesses as a function of $\nu$. To finish modifying our proofs to the general setting -- it is enough to prove an analogous condition to the above for the Bayesian DM, which is provided by the following~lemma.\footnote{Note that the expected utility of the ignorant DM in both settings is the convex combination of the two quantities above, with respective weights $\mu$ and $1-\mu$.}
\begin{lemma}
\label{lem:states}
Fix $\omega'\in\Omega$. Then for any information structure $\pi$ in the general setting and the corresponding induced information structure $\pi'$ in the binary-posterior setting from Section~\ref{sec:preliminaries}, it holds that:
\begin{align*}
   \mathbb{E}_{\nu'\sim\hat{\pi'}_{\omega'}}\brackets*{\Pr\brackets*{s^B\parentheses*{\nu'}\neq \omega'}}=\mathbb{E}_{\nu\sim\hat{\pi}_{\omega'}}\brackets*{\Pr\brackets*{s^B\parentheses*{\nu}\neq \omega'}}
\end{align*}
\end{lemma}
\begin{proof}
The key idea is that the Bayesian DM cannot decrease the mistake probability knowing $\pi$ compared to $\pi'$ as she does not observe the exact posterior of each agent, but only her recommendation. We shall only prove the lemma for $\omega'=1$ (the case $\omega'=0$ is analogous). Indeed, we have:
\begin{align*}
    &\mathbb{E}_{\nu'\sim\hat{\pi'}_1}\brackets*{\Pr\brackets*{s^B\parentheses*{\nu'}\neq 1}}=\Pr_{q\sim\pi'\parentheses*{\omega=1}}\brackets*{q<1/2}=\Pr_{q\sim\pi'\parentheses*{\omega=1}}\brackets*{q=p_1}=_{\parentheses*{*}}\\
    &\Pr_{q\sim\pi\parentheses*{\omega=1}}\brackets*{q<1/2}=\mathbb{E}_{\nu\sim\hat{\pi}_1}\brackets*{\Pr\brackets*{s^B\parentheses*{\nu}\neq 1}},
\end{align*}
where $\parentheses*{*}$ holds by definition of $\pi'$.
\end{proof}
\subsection{Arbitrary State Space and Utilities}
\label{sub:states}
In this section, we weaken the assumption that the state space is binary and the DM just tries to guess the correct state. Rather, we just assume that 
$\Omega$ is finite, and the state space is equipped with some prior distribution $\mu=\parentheses*{\mu_{\omega}}_{\omega\in\Omega}$. The DM has to choose between a \emph{default action} yielding her a constant utility of $0$ regardless of the state, and an \emph{optional action} yielding her a state-dependent utility according to some function $u:\Omega\to\mathbb{R}$.\footnote{Note that the normalization of the DM's utility in one of the actions to be constantly $0$ is w.l.o.g.}

Each agent recommends taking one of the actions. An agent with a posterior belief $q\in\Delta\parentheses*{\Omega}$ recommends taking the optional action if, in expectation over $q$, the DM is weakly better off taking it. Formally: $\mathbb{E}_{\omega\sim q}\brackets*{u\parentheses*{\omega}}\geq 0$. Otherwise, she recommends taking the default action. 
The (marginal) information structure $\pi_i$ for a single agent is identical for all agents and is given as follows. Agent $i$ ($i\in\brackets*{n}$) gets a binary private signal $s_i\in\braces*{L,H}=S$, with the corresponding posteriors being $p^L=\parentheses*{p^L_{\omega}}_{\omega\in\Omega},p^H=\parentheses*{p^H_{\omega}}_{\omega\in\Omega}$. We shall assume that $\mathbb{E}_{\omega\sim p^H}\brackets*{u\parentheses*{\omega}}> 0$ and $\mathbb{E}_{\omega\sim p^L}\brackets*{u\parentheses*{\omega}}<0$ -- that is, upon observing $H$ the agent would recommend taking the optional action, while upon observing $L$ she would recommend the default action.\footnote{If an agent would (weakly) recommend the same action regardless of the signal, then the DM can just ignore the recommendations.} Denote by $\Omega^H\subseteq\Omega$ the set of states for which the DM is weakly better off taking the optional action, and define $\Omega^L:=\Omega\setminus\Omega^H$. That is, $\Omega^H=\braces*{\omega'\in\Omega: u\parentheses*{\omega'}\geq 0}$ and $\Omega^L=\braces*{\omega'\in\Omega: u\parentheses*{\omega'}< 0}$.

The ignorant DM and the Bayesian DM both observe the fraction of agents recommending $H$. The ignorant DM knows just the identical marginal information structure $\pi_i\in\Delta\parentheses*{\Omega\times S}$, while the Bayesian DM is aware of the joint information structure $\pi\in\Delta\parentheses*{\Omega\times S^n}$. The aggregation rule $f$ gets as input the fraction of agents recommending $H$ and outputs the probability of taking the optional~action.

For any state $\omega'\in\Omega$, the probability of a specific agent to have the high posterior $p^H$ conditional on the state $\omega'$ is:
\begin{equation*}
a_{\omega'}=\frac{p^H_{\omega'}\parentheses*{\mu_{\omega'}-p^L_{\omega'}}}{\mu_{\omega'}\parentheses*{p^H_{\omega'}-p^L_{\omega'}}}.  
\end{equation*}
The regret of an aggregation rule $f$, $\reg\parentheses*{f}$, is the difference between the expected utilities of a Bayesian DM knowing $\pi$ and the ignorant DM not knowing $\pi$. The following generalization of Theorem~\ref{thm:dictator} holds.
\begin{theorem}
\label{thm:arb}
Suppose $\frac{1}{n}\leq a_{\omega'} \leq \frac{n-1}{n}$ for every $\omega'\in\Omega$. Then the random dictator aggregation rule is uniquely optimal. Moreover:
\begin{align*}
  &\reg=\sum_{\omega'\in\Omega^H} \mu_{\omega'}u\parentheses*{\omega'}-\sum_{\omega'\in\Omega} \mu_{\omega'} a_{\omega'} u\parentheses*{\omega'}.
\end{align*}
\end{theorem}
Let us describe the intuition behind the proof of Theorem~\ref{thm:ab_gen}, which follows similar ideas to the proof of Theorem \ref{thm:ab} (the binary case). We first note that the set of feasible distributions over frequencies of posteriors has an analogous neat characterization for an arbitrary state space (rather than binary).
\begin{theorem}[\citet{arieli2022population} - A generalization of Theorem~\ref{thm:ab}]\label{thm:ab_gen}
Let  $\parentheses*{\hat{\rho}_{\omega'}}_{\omega'\in\Omega}\in\parentheses*{\Delta\parentheses*{\braces*{0,\frac{1}{n},\ldots,1}}}^{\absolute*{\Omega}}$ be a tuple of conditional distributions over frequencies of the $H$-signal. There exists an information structure $\pi$ with $\hat{\pi}_{\omega'}=\hat{\rho}_{\omega'}$ for every $\omega'\in\Omega$ if and only if $\mathbb{E}\brackets*{\hat{\rho}_{\omega'}}=a_{\omega'}$ for every $\omega'\in\Omega$.
\end{theorem}
Following the same outline as of Proposition~\ref{pro:concav} proof, with the use of Theorem~\ref{thm:ab} replaced by Theorem~\ref{thm:ab_gen}, we deduce the following result (the proof is omitted due to the similarity to Proposition~\ref{pro:concav}).
\begin{proposition}
\label{pro:concav_gen}
For every $\braces*{a_{\omega'}}_{\omega'\in\Omega},n,\mu,$ and $f$:
\begin{align*}
    &\reg \parentheses*{f}\leq \sum_{\omega'\in\Omega^H} \mu_{\omega'} u\parentheses*{\omega'}-\sum_{\omega'\in\Omega^L}\mu_{\omega'}\cav\brackets*{f}\parentheses*{a_{\omega'}}u\parentheses*{\omega'}-\sum_{\omega'\in\Omega^H}\mu_{\omega'}\vex\brackets*{f}\parentheses*{a_{\omega'}}u\parentheses*{\omega'}.
\end{align*}
\end{proposition}
Similarly to the binary-state case, whenever there exists an optimal $\pi$ for which the supports of $\hat{\pi_{\omega'}}$ with $\omega'\in\Omega^H$ are disjoint from the supports of $\hat{\pi_{\omega'}}$ with $\omega'\in\Omega^L$ -- the inequality in Proposition~\ref{pro:concav_gen} is tight. Turning this idea into Theorem~\ref{thm:ab} proof is similar to Theorem~\ref{thm:dictator} proof, and it is presented in Appendix~\ref{app:arb}.

As a corollary of Theorem \ref{thm:arb}, given any fixed prior and marginal posteriors -- for a large enough number of agents $n$, the random dictator rule is uniquely optimal. One can further drop the assumption that each agent gets a binary signal (as was done in Subsection~\ref{sub:posteriors}).
\subsection{Non-Anonymity}
\label{sub:anonymity}
We now prove that the assumption that the DM does not observe the identities of the agents is w.l.o.g. Namely, let us consider a setting in which the DM's aggregation rule $f:2^{\brackets*{n}}\to\brackets*{0,1}$ depends also on the identity of the agents, where the domain of $f$ is the set of all agents getting the high signal.\footnote{Note that as in the previous two subsections, one can extend the results to arbitrary posteriors, and extend Theorem~\ref{thm:dictator} and its corollaries to arbitrary finite state spaces.} Denote by $\widehat{P^*}\parentheses*{\pi}, \widehat{P}\parentheses*{f,\pi},\widehat{\reg}\parentheses*{f,\pi}, \widehat{\reg}\parentheses*{f}$ and $\widehat{\reg}$ the following quantities in the non-anonymous setting, respectively: the probability of a correct guess by the Bayesian DM under the information structure $\pi$, the probability of a correct guess by the ignorant DM under the information structure $\pi$ and the aggregation rule $f$, the regret under $f$ and $\pi$, the maximum regret of $f$ and the minimal regret in the setting. We claim that $\reg=\widehat{\reg}$. 

For a permutation $\tau$ on $\brackets*{n}$, an aggregation rule $f$ and a distribution over information structures $\phi$, let $f_{\tau}$ and $\phi_{\tau}$ be, respectively, the aggregation rule and the information structure obtained from $f$ and $\phi$ upon applying $\tau$ on the agents' identities. We need the following~lemma.
\begin{lemma}
    \label{lem:anonymity}
    Consider the two-player zero-sum game between the DM and the adversary in which the DM chooses a non-anonymous aggregation rule $f$, the adversary chooses a distribution over information structures $\phi$, and the payoff is $\mathbb{E}_{\pi\sim\phi}\brackets*{\widehat{\reg}\parentheses*{f,\pi}}$. Then there exist optimal DM's and adversary's strategies $f^*$ and $\phi^*$, respectively, such that $f^*=f^*_{\tau}$ and $\phi^*=\phi^*_{\tau}$ for any permutation $\tau$ on $\brackets*{n}$.
\end{lemma}
Note that Lemma~\ref{lem:anonymity} immediately implies $\reg=\widehat{\reg}$. Indeed, let $f^*$ and $\phi^*$ be non-anonymous optimal aggregation rule and distribution over information structures (respectively) invariant under agents' permutations. If $f^*$ is suboptimal in the anonymous setting, then there exists $\phi$ with $\mathbb{E}_{\pi\sim\phi}\brackets*{\reg\parentheses*{f^*,\pi}}>\mathbb{E}_{\pi\sim\phi^*}\brackets*{\reg\parentheses*{f^*,\pi}}$. However, we have $\mathbb{E}_{\pi\sim\phi}\brackets*{\reg\parentheses*{f^*,\pi}}=\mathbb{E}_{\pi\sim\phi}\brackets*{\widehat{\reg}\parentheses*{f^*,\pi}}$ and $\mathbb{E}_{\pi\sim\phi^*}\brackets*{\reg\parentheses*{f^*,\pi}}=\mathbb{E}_{\pi\sim\phi^*}\brackets*{\widehat{\reg}\parentheses*{f^*,\pi}}$, implying $\mathbb{E}_{\pi\sim\phi}\brackets*{\widehat{\reg}\parentheses*{f^*,\pi}}>\mathbb{E}_{\pi\sim\phi^*}\brackets*{\widehat{\reg}\parentheses*{f^*,\pi}}$, a contradiction. Similarly, we get that $\phi^*$ is optimal in the anonymous setting.

\begin{proof}[Proof of Lemma~\ref{lem:anonymity}]
Fix an optimal aggregation rule $f$. Define $f^*:2^{\brackets*{n}}\to\brackets*{0,1}$ by $f^*\parentheses*{S}:=\sum_{\tau}\frac{f_{\tau}\parentheses*{S}}{n !}$. Then trivially $f^*\parentheses*{S}=\frac{\sum_{S'\subseteq \brackets*{n}: \absolute*{S'}=\absolute*{S}} f\parentheses*{S}}{\binom{n}{\absolute*{S}}}$ depends only on $\absolute*{S}$ -- i.e., it is invariant under permutations on the agents' identities. Let $\phi$ be an adversary's optimal strategy. By the agents' symmetry, the aggregation rule $f_{\tau}$ is optimal for any permutation $\tau$ on $\brackets*{n}$. Thus, $\widehat{\reg}=\mathbb{E}_{\pi\sim\phi}\brackets*{\widehat{\reg}\parentheses*{f_{\tau},\pi}}$ for any $\tau$, implying $\widehat{\reg}=\mathbb{E}_{\pi\sim\phi}\brackets*{\widehat{\reg}\parentheses*{f^*,\pi}}$. Therefore, $f^*$ is an optimal aggregation rule invariant under agents' permutations.

Define $\phi^*$ to be a distribution over information structures choosing uniformly at random a permutation $\tau$ on $\brackets*{n}$ and then choosing an information structure according to $\phi_{\tau}$. Since $f^*$ is invariant under any permutation $\tau$ on the agents, and the agents are symmetric -- we get $\mathbb{E}_{\pi\sim\phi_{\tau}}\brackets*{\widehat{\reg}\parentheses*{f^*,\pi}}=\mathbb{E}_{\pi\sim\phi}\brackets*{\widehat{\reg}\parentheses*{f^*,\pi}}$ for any permutation $\tau$, implying $\mathbb{E}_{\pi\sim\phi^*}\brackets*{\widehat{\reg}\parentheses*{f^*,\pi}}=\mathbb{E}_{\pi\sim\phi}\brackets*{\widehat{\reg}\parentheses*{f^*,\pi}}=\widehat{\reg}$. Thus, $\phi^*$ is an optimal adversary's strategy invariant under agents' permutations.
\end{proof}
\subsection{Approximation Ratio}
\label{sub:approx}
In this section, we consider an alternative robustness paradigm -- \emph{approximation ratio}. This approach differs from regret analysis by considering the \emph{ratio} of the probabilities of the correct guesses of the ignorant and the Bayesian DM, rather than their difference. Formally, the \emph{approximation ratio of the aggregation rule $f$ in an information structure $\pi$} is $\appr\parentheses*{f,\pi}:=\frac{P\parentheses*{f,\pi}}{P^*\parentheses*{\pi}}$. The \emph{approximation ratio of an aggregation rule $f$} is $\appr\parentheses*{f}:=\min_{\pi} \appr\parentheses*{f,\pi}$.\footnote{Note that this minimum exists since $\appr\parentheses*{f,\cdot}$ is a continuous functional defined on a compact~space.}
\begin{proposition}
    \label{pro:approx}
    Suppose $\frac{1}{n}\leq a < b \leq \frac{n-1}{n}$.\footnote{Note that this condition holds for all $n\geq \max\braces*{\frac{\parentheses*{1-\mu}\parentheses*{p_2-p_1}}{\parentheses*{1-p_2}\parentheses*{\mu-p_1}},\frac{\mu\parentheses*{p_2-p_1}}{p_1\parentheses*{p_2-\mu}}}$.} Then random dictator is the unique approximation ratio-maximizing aggregation rule. Moreover,
$\appr=\parentheses*{1-\mu}\cdot\parentheses*{1-a}+\mu\cdot b$.
\end{proposition}
\begin{proof}
    Let $f$ be an aggregation rule. By the proof of Theorem~\ref{thm:dictator}, for $\frac{1}{n}\leq a < b \leq \frac{n-1}{n}$ -- when the adversary targets maximizing the regret, there always exists an optimal adversary's information structure $\pi$ for which the supports of $\hat{\pi_0}$ and $\hat{\pi_0}$ are disjoint. Such strategy reveals the value of $\omega$ to a Bayesian DM with the maximum possible probability of $1$. Thus, $\appr\parentheses*{f}$ is at most the probability of a correct guess of $\omega$ under $f$, which is shown in Proposition~\ref{pro:minimax} to be at most $\parentheses*{1-\mu}\cdot\parentheses*{1-a}+\mu\cdot b$, with the bound achieved exclusively by the random dictator rule.

    Conversely, let the DM use the random dictator rule. Then the proof of Proposition~\ref{pro:minimax}  implies that the DM guesses $\omega$ correctly with probability $\parentheses*{1-\mu}\cdot\parentheses*{1-a}+\mu\cdot b$. In particular, the approximation ratio of the random dictator is lower-bounded by this value, as needed.
\end{proof}
We should also mention that one can get an analogous result in the setting with arbitrary state space and utilities presented in Subsection~\ref{sub:states} -- the random dictator rule remains~optimal.
\subsection{Strategic Agents}
\label{sub:str}
We have assumed that agents simply report the better recommendation conditional on their information. Our robustness analysis advocates the usage of the random dictator rule even for strategic agents. Here we highlight another desirable property of the random dictator rule: If agents are strategic and they know that the aggregation rule is the random dictator, then it is a dominant strategy for them to report the better recommendation conditional on their information. Namely, in addition to being robustly optimal, the random dictator is also immune to scenarios in which some (or all) agents start acting strategically instead of truthfully.
\section{Conclusions and Future Work}
\label{sec:conclusion}
We initiate the study of regret minimization for aggregating information. While the regret minimization approach turns out to be fundamentally different from the minimax approach for a small number of symmetric agents -- with the former being much more intricate -- they are equivalent for a sufficiently large number of such agents. Moreover, the approximation ratio robustness approach is also equivalent to these approaches when the number of agents is large.

Our work suggests many natural open questions. The tractability of our analysis crucially relies on the fact that the decision of the DM is binary. It remains an open problem how to analyze settings with more than two actions for the DM. Weakening the assumption that the agents are symmetric is another interesting direction for the regret approach. Furthermore, an online variant of our setting -- with the DM trying to learn the optimal aggregation rule -- remains unexplored. Another interesting challenge is to come up with mild limitations on the power of the adversary under which natural aggregation rules -- such as supermajority -- perform better than the random dictator and potentially are robustly optimal. Finally, the existence of a \emph{universally} robust strategy (aggregation rule in our case) that is optimal under several robustness approaches is a seemingly rare phenomenon, which we have found to hold in the information aggregation problem. We do not know of any other setting in which such a universally robust strategy exists. It will be interesting to find other settings with this phenomenon and explore underlying reasons.

\paragraph{Acknowledgements.} 
The authors are grateful to Kfir Eliaz, Sergiu Hart and anonymous reviewers for their helpful remarks, including suggesting the paper's title. Itai Arieli is supported by the Ministry of Science and Technology grant No.~2028255. Yakov Babichenko is supported by the Binational Science Foundation BSF grant No.~2018397 and by the German-Israeli Foundation for Scientific Research and Development GIF grant No.~I-2526-407.6/2019. This work is funded by the European Union (ERC, ALGOCONTRACT, 101077862, PI: Inbal Talgam-Cohen). Inbal Talgam-Cohen is supported by a Google Research Scholar award and by the Israel Science Foundation grant No.~336/18. Yakov and Inbal are supported by the Binational Science Foundation BSF grant No.~2021680. Konstantin Zabarnyi is supported by a PBC scholarship for Ph.D. students in data science. 

\bibliographystyle{plainnat}
\begin{singlespacing}
\bibliography{main}
\end{singlespacing}

\appendix
\section{Minimax versus Regret}
\label{ap:vs}
In this appendix, we demonstrate that the celebrated result of~\citet{carroll2017robustness} on the minimax-optimality of selling goods separately does not extend to regret optimality. We further discuss another setting in which different notions of robustness lead to different robustly optimal strategies~\cite{babichenko2022regret}.

\citet{carroll2017robustness}~considers a classical combinatorial auction setting. A single seller suggests $n$ goods to a single buyer. The buyer's valuation for the goods is additive, and it is described by a vector $v=\parentheses*{v_i}_{i\in \brackets*{n}}$ that specifies her utility for each good. The vector $v$ is drawn according to a distribution $\pi$. By the direct revelation principle, a mechanism for selling the goods might be described as a (possibly infinite) menu $M$ from which the buyer picks a single item. Each menu item takes the form $m=\parentheses*{q_1,...,q_n,p}\in M$, where $q_i\in \brackets*{0,1}$ specifies the probability of the good to be allocated to the buyer, and $p>0$ is the price of this menu item. We denote by $M^*\parentheses*{\pi}$ the optimal mechanism (menu). Let $s\parentheses*{M,\pi}$ denote the seller's revenue in case she uses mechanism $M$.~\citet{carroll2017robustness}~studies a scenario in which the seller does not know $\pi$ (and hence cannot pick $M^*\parentheses*{\pi}$), but only knows the marginal distribution of valuations on each good separately. Namely, the seller knows $\pi_i$, where $\pi_i$ is the distribution of $v_i$. The correlation between the $v_i$s is unknown to the seller. The seller chooses a mechanism $M$ as a function of $(\parentheses*{\pi_1,...,\pi_n}$ only. The seller's goal is to choose a mechanism that performs robustly well; namely, for all possible correlations. Therefore, the \emph{minimax} guarantee and the \emph{regret} guarantee of a mechanism $M$ are defined by:
\begin{align*}
    &\minmax\parentheses*{M}=\inf_{\pi \text{ with marginals } \parentheses*{\pi_1,...,\pi_n}} s\parentheses*{M,\pi}, \\
    &\reg\parentheses*{M}=\sup_{\pi \text{ with marginals } \parentheses*{\pi_1,...,\pi_n}} \brackets*{s\parentheses*{M^*\parentheses*{\pi},\pi}-s\parentheses*{M,\pi}}.
\end{align*}
The result of~\cite{carroll2017robustness} states that selling the goods separately (with an optimal price on each good $i$ determined by $\pi_i$ only), which is denoted by $M_{sep}$, is minimax-optimal -- i.e., it maximizes $\minmax\parentheses*{M}$ across all mechanisms). Below we demonstrate an example in which $M_{sep}$ is not regret-optimal (i.e., it does not minimize $\reg\parentheses*{M}$).

Assume that $n$ is even and $\pi_i$ is the uniform distribution over $\braces*{1,2}$ for all goods $i\in \brackets*{n}$. An optimal price for each good separately is $1$, and it ensures a deterministic revenue of $1$. Hence, the total revenue is $n$. We argue that $\reg\parentheses*{M_{sep}}=\frac{n}{2}$. Indeed, let $\pi$ be the uniform distribution over the two vectors $\braces*{v,v'}$, where $v=\parentheses*{1,2,1,2,\ldots,1,2}$ and $v'=\parentheses*{2,1,2,1,\ldots,2,1}$. Selling all goods as a single bundle for $\frac{3n}{2}$, which is denoted by $M_{bun}$, ensures selling with probability $1$, yielding a revenue of $\frac{3n}{2}$ for the seller. Moreover, the seller cannot extract more than $\frac{3n}{2}$ revenue, as the revenue is bounded by the buyer's expected value for the entire bundle.

Assume by way of contradiction that $M_{sep}$ minimizes regret; i.e., the minimal regret is $\frac{n}{2}$. We view the regret-minimization problem as a zero-sum game between the seller (who chooses $M$) and the adversary (who chooses $\pi$). A (mixed) strategy of the adversary in this zero-sum game is a distribution $\phi$ over $\pi$s. By the minimax theorem, there exists a mixed strategy $\phi$ for the adversary such that for every $M$, we have: $\mathbb{E}_{\pi \sim \phi} \brackets*{s\parentheses*{M^*\parentheses*{\pi},\pi}-s\parentheses*{M,\pi}} \geq \frac{n}{2}$. In particular, this should hold for $M=M_{sep}$. Note, however, that $s\parentheses*{M^*\parentheses*{\pi},\pi}\leq \frac{3n}{2}$ (because it is buyer's expected value for the entire bundle) and that $s(M_{sep},\pi)=n$. Therefore, to achieve the $\frac{n}{2}$ difference $\phi$ must be supported on $\pi$s for which $s\parentheses*{M^*\parentheses*{\pi},\pi}=\frac{3}{2}$. This can happen only if all goods are sold with probability 1 (otherwise the full revenue cannot be extracted) and we must have $v_1+...+v_n=\frac{3n}{2}$ with probability 1 (again because otherwise the full revenue cannot be extracted). Therefore, $\phi$ is supported on $\pi$s for which the buyer's value for the entire bundle is $\frac{3n}{2}$ with probability 1. Therefore, the seller can use $M_{bun}$ against $\phi$ and have a regret of $0$, which is a contradiction. 

\paragraph{Further discussion.} 
Another example demonstrating the differences between robustness approaches appears in~\citet{babichenko2022regret}. \citet{babichenko2022regret}~study the Bayesian persuasion problem~\cite{kamenica2011bayesian} with an uncertainty of the sender about the receiver's utility. In this setting, the minimax approach is very pessimistic: It is hopeless to have a non-trivial upper bound when the number of states is large. The regret approach succeeds in providing non-trivial asymptotic (in the number of states) bounds on the regret. Interestingly, in the same setting providing non-trivial asymptotic bounds on the approximation ratio turns out to be impossible, whereas, for a constant number of states, bounds are provided.
\section{Proof of Theorem~\ref{thm:dictator}}
\label{app:dictator}
We first lower-bound $\reg$ under the assumption that one can separate the supports of $\hat{\pi}_0$ and $\hat{\pi}_1$. We shall show that the DM betters off by replacing $f$ with the segment connecting the points $\parentheses*{0,f\parentheses*{0}}$ and $\parentheses*{1,f\parentheses*{1}}$.
\begin{lemma}
\label{lem:lower}
Fix $f$ and let $l$ be the line passing through $\parentheses*{0,f\parentheses*{0}}$ and $\parentheses*{1,f\parentheses*{1}}$. Suppose there exist distinct $x_1,x_2,x_3,x_4\in\braces*{0,\frac{1}{n},\ldots,1}$ s.t.~$x_1\leq a\leq x_2$, $x_3\leq b\leq x_4$ with $\parentheses*{x_1,f\parentheses*{x_1}},\parentheses*{x_2,f\parentheses*{x_2}}$ not below $l$ and $\parentheses*{x_3,f\parentheses*{x_3}},\parentheses*{x_4,f\parentheses*{x_4}}$ not above $l$.\footnote{Formally, e.g., $f\parentheses*{x_1}\geq \parentheses*{f\parentheses*{1}-f\parentheses*{0}}\parentheses*{x-x_1}+f\parentheses*{0}$.} Then:
\begin{align*}
    &\reg\parentheses*{f}\geq \reg\parentheses*{l}=1-\parentheses*{1-\mu}\cdot\parentheses*{1- a\parentheses*{f\parentheses*{1}-f\parentheses*{0}}}-\mu\parentheses*{b\parentheses*{f\parentheses*{1}-f\parentheses*{0}}}-\parentheses*{2\mu-1}f\parentheses*{0}.
\end{align*}
\end{lemma}
\begin{proof}
Assume that there exist $x_1,x_2,x_3,x_4$ as described. Note first that since $l$ is a linear function, $P\parentheses*{l,\pi}$ is independent of the information structure $\pi$. Moreover, $P^*\parentheses*{\pi}=1$ for $\pi$ inducing the distributions $\hat{\pi}_0,\hat{\pi}_1$ over values of $\nu$ with $\supp\parentheses*{\hat{\pi}_0}=\braces*{x_1,x_2}$ and $\supp\parentheses*{\hat{\pi}_1}=\braces*{x_3,x_4}$, and respective expectations $a,b$ (note that the distributions $\hat{\pi}_0,\;\hat{\pi}_1$ are uniquely determined). Indeed, separating these supports allows the Bayesian DM to guess $\omega$ correctly with probability $1$. Thus, $\reg\parentheses*{l}=1-P\parentheses*{l,\pi}$ for the above $\pi$. As $\reg\parentheses*{f}\geq \reg\parentheses*{f,\pi}$, it is enough to prove that $P\parentheses*{f,\pi}\leq P\parentheses*{l,\pi}$.

Indeed, by the assumption on $x_1,x_2,x_3,x_4$, we have:\\ $\mathbb{E}_{X_0\sim\hat{\pi}_0}\brackets*{1-f\parentheses*{X_0}} \leq\parentheses*{1- a\parentheses*{f\parentheses*{1}-f\parentheses*{0}}}-f\parentheses*{0}$. Similarly, we get:\\$\mathbb{E}_{X_1\sim\hat{\pi}_1}\brackets*{f\parentheses*{X_1}} \leq b\parentheses*{f\parentheses*{1}-f\parentheses*{0}}+f\parentheses*{0}$. Therefore:
\begin{align*}
&P\parentheses*{f,\pi}\leq\parentheses*{1-\mu}\cdot\parentheses*{1- a\parentheses*{f\parentheses*{1}-f\parentheses*{0}}-f\parentheses*{0}}+\mu\parentheses*{b\parentheses*{f\parentheses*{1}-f\parentheses*{0}}+f\parentheses*{0}}=P\parentheses*{l,\pi},
\end{align*}
and the lemma statement follows.
\end{proof}
Now we are ready to prove Theorem~\ref{thm:dictator}. We first show, using Lemma~\ref{lem:lower}, that there always exists an optimal $f$ which is a linear function; the required $x_1,x_2,x_3,x_4$ from Lemma~\ref{lem:lower} statement would be chosen from the set $\braces*{0,\frac{1}{n},1-\frac{1}{n},1}$. Then we prove that the optimal $f$ must be the random dictator.
\begin{proof}[Proof of Theorem~\ref{thm:dictator}]
Let $l$ be the line passing through $\parentheses*{0,f\parentheses*{0}}$ and $\parentheses*{1,f\parentheses*{1}}$. We shall first prove that for any choice of $f$, the inequality ensured by Lemma~\ref{lem:lower} holds. To find $x_1,x_2,x_3,x_4$ as needed in Lemma~\ref{lem:lower}, consider the possible cases regarding the graph of $f$:
\begin{enumerate}
    \item $\parentheses*{\frac{1}{n},f\parentheses*{\frac{1}{n}}}$ is not above $l$ and $\parentheses*{\frac{n-1}{n},f\parentheses*{\frac{n-1}{n}}}$ is not below $l$. Then one may take $x_1=0,\; x_2=\frac{n-1}{n},\; x_3=\frac{1}{n},\; x_4=1$ (note that $n>2$).
    \item $\parentheses*{\frac{1}{n},f\parentheses*{\frac{1}{n}}}$ is strictly above $l$, and $\parentheses*{\frac{n-1}{n},f\parentheses*{\frac{n-1}{n}}}$ is not below $l$. This time, one can take $x_1=\frac{1}{n},\; x_2=\frac{n-1}{n},\; x_3=0,\; x_4=1$.
    \item $\parentheses*{\frac{n-1}{n},f\parentheses*{\frac{n-1}{n}}}$ is strictly below $l$, and $\parentheses*{\frac{1}{n},f\parentheses*{\frac{1}{n}}}$ is not above $l$. This case is symmetric to the previous one.
    \item Otherwise, $\parentheses*{\frac{1}{n},f\parentheses*{\frac{1}{n}}}$ is strictly above $l$ and $\parentheses*{\frac{n-1}{n},f\parentheses*{\frac{n-1}{n}}}$ is strictly below $l$. One can take $x_1=\frac{1}{n},\; x_2=1,\; x_3=0,\; x_4=\frac{n-1}{n}$.
\end{enumerate}

Therefore, there exists an optimal $f$ which is linear -- i.e., $f$ coincides with $l$. Fix an optimal linear $f$ and write $f\parentheses*{x}=mx+k$ for some constants $m,k$. As $k=f\parentheses*{0}$ and $m=f\parentheses*{1}-f\parentheses*{0}$, we get:
\begin{align*}
    &\reg\parentheses*{f}= 1 - \parentheses*{\parentheses*{1-\mu}\cdot\parentheses*{1- a\parentheses*{f\parentheses*{1}-f\parentheses*{0}}}+\mu\parentheses*{b\parentheses*{f\parentheses*{1}-f\parentheses*{0}}}+\parentheses*{2\mu-1}f\parentheses*{0}}=\\
    &\mu+k\cdot \parentheses*{1-2\mu}+m\cdot\parentheses*{\parentheses*{1-\mu}a-\mu b}.
\end{align*}
The only constraints on $m,k$ are $k=f\parentheses*{0}\in \brackets*{0,1}$ and $m+k=f\parentheses*{1}\in \brackets*{0,1}$. Note that $\parentheses*{1-\mu}a-\mu b=\frac{\parentheses*{\mu-p_1}\parentheses*{1-2p_2}}{p_2-p_1}<1-2\mu$ as $p_1<\frac{1}{2},\mu<p_2$.
Therefore, to minimize $\reg\parentheses*{f}$, one should take $k=0$. Moreover, as $p_2>\frac{1}{2}$, one should take $m=1$. Thus, $\reg\parentheses*{f}\geq 1-\parentheses*{1-\mu}\parentheses*{1-a}-\mu b$. Since the right-hand side is exactly the regret of the random dictator aggregation rule, the optimality of the random dictator follows.

To show uniqueness, consider some optimal $f$. If $f$ is not the random dictator rule, there exists some $i\in\brackets*{n}$ with $\parentheses*{\frac{i}{n},f\parentheses*{\frac{i}{n}}}$ either strictly above $l$ or strictly below $l$. Assume w.l.o.g. That the former holds. The inequality in Lemma~\ref{lem:lower} must be tight and the lower bound provided by this lemma must equal $1-\parentheses*{1-\mu}\parentheses*{1-a}-\mu b$. Thus, we must have that for any choices of $x_1,x_2,x_3,x_4$ satisfying the conditions of Lemma~\ref{lem:lower}: $i\neq x_1,x_2,x_3,x_4$. In particular, $1<i<n-1$. Note that either $\frac{i}{n}\leq a$ or $\frac{i}{n}\geq a$. Assume w.l.o.g.~that the former holds. Then re-selecting $x_1$ form the value in $\braces*{0,\frac{1}{n}}$ in Lemma~\ref{lem:lower} proof to $x_1=\frac{i}{n}$ prevents from the inequality in Lemma~\ref{lem:lower} to be tight, as the restriction of $f$ to $x_1,x_2,x_3,x_4$ is not linear.
\end{proof}
\section{Analysis of Example~\ref{ex:two}}
\label{app:two}
\begin{proof}
Similarly to the proof of Theorem \ref{thm:gap}, we consider a two-player zero-sum game between the DM and the adversary. The DM's pure strategies are choices of an aggregation function $f:\braces*{0,\frac{1}{n},\ldots,1}\to \braces*{0,1}$ and the adversary's pure strategies are choices of $x\in C$ (representing admissible information structures). Denote the adversary's mixed strategy by $\phi$; note that mixtures over admissible information structures are themselves admissible information structures. The payoff function is:
\begin{align*}
     &\reg\parentheses*{f,\phi}=\mathbb{E}_{x\sim\phi}\Bigg[\sum_{i=0}^2\left(\max\braces*{ \sum_{D\subseteq\braces*{0,1}:\;\absolute*{D}=i} x^0_D, \sum_{D\subseteq\braces*{0,1}:\;\absolute*{D}=i} x^1_D}\right. \nonumber\\
     &\qquad \left.-\parentheses*{\parentheses*{1-f_i}\cdot \sum_{D\subseteq\braces*{0,1}:\;\absolute*{D}=i} X_D^0 +f_i \cdot \sum_{D\subseteq\braces*{0,1}:\;\absolute*{D}=i} x_D^1}\right)\Bigg].
\end{align*}
Denote the function $f$ described in the proposition statement by $f^*$. We shall show that there exists $\phi^*$ s.t.~$\parentheses*{f^*,\phi^*}$ is a mixed Nash equilibrium, which would complete the proof (after further computing the game value). We divide the proof into three cases according to the order between $a,b$, and $1/2$.

\paragraph{{\bf Case 1: $a<b\leq 1/2$.}} Each $x\in\supp\parentheses*{\phi}$ that is a best reply to $f$ satisfying $f\parentheses*{0}=0$, $f\parentheses*{1}=1$ must be a solution of the following convex program:\footnote{We denote $p:=x_{1,2}^0$, $q:=x_{1,2}^1$ and use the definition of $C$.}
\begin{align*}
    &\max_{p\in\brackets*{0,a/2},\;q\in\brackets*{0,b/2}} \biggl\{\max\braces*{1/2-a+p,1/2-b+q}+2\max\braces*{a/2-p,b/2-q}+\max\braces*{p,q}\\
    &-\parentheses*{1/2-a+p}-2\parentheses*{a/2-p}\parentheses*{1-f\parentheses*{1/2}}-2\parentheses*{b/2-q}f\parentheses*{1/2}-q\biggr\}.
\end{align*}
Denote the above target function by $h\parentheses*{p,q}$. As $h$ is a convex function -- it obtains its maximum at an extreme point of the domain. Hence, the only candidates for solutions are $\parentheses*{p,q}\in\braces*{\parentheses*{0,0},\parentheses*{a/2,0},\parentheses*{0,b/2},\parentheses*{a/2,b/2}}$. Note that:
\begin{itemize}
    \item $h\parentheses*{0,0}=\parentheses*{b-a}\parentheses*{1-f\parentheses*{1/2}}$.
    \item $h\parentheses*{a/2,0}=a/2+b\parentheses*{1-f\parentheses*{1/2}}>_{\text{For }a>0} h\parentheses*{0,0}$.
    \item $h\parentheses*{0,b/2}=\max\braces*{1/2-a,1/2-b/2}+a-1/2+a f\parentheses*{1/2}$.
    \item $h\parentheses*{a/2,b/2}=0<_{\text{For }a>0} h\parentheses*{a/2,0}$.
\end{itemize}
Therefore, the only remaining candidates for maximizers of $h$ are $\parentheses*{a/2,0}$ and $\parentheses*{0,b/2}$, regardless of $f$.  A straightforward check implies that for the $f$ minimizing $\max\braces*{h\parentheses*{a/2,0},h\parentheses*{0,b/2}}$, it must hold that $h\parentheses*{a/2,0}=h\parentheses*{0,b/2}$. If $2a\leq b$, we get $f\parentheses*{1/2}=\frac{a/2+b}{a+b}=\frac{a+2b}{2\parentheses*{a+b}}$ and $\reg=\frac{a\parentheses*{a+2b}}{2\parentheses*{a+b}}$. Otherwise, we get $f\parentheses*{1/2}=\frac{3b-a}{2\parentheses*{a+b}}$ and $\reg=\frac{a^2+4ab-b^2}{2\parentheses*{a+b}}$.

Write $x\in C$ as $x=\parentheses*{x_{\emptyset}^0,x_{1}^0,x_{2}^0,x_{1,2}^0,x_{\emptyset}^1,x_{1}^1,x_{2}^1,x_{1,2}^1}$. Given an adversary's (pure) strategy $x\in C$, let $\delta_x$ be the embedding of $x$ into the space of adversary's mixed strategies; that is, $\delta_x$ is a distribution over elements of $C$ assigning probability $1$ to $x$. From the arguments above, it follows that any distribution of the form: $\phi^*=\alpha\delta_{x^a}+\parentheses*{1-\alpha}\delta_{x^b}$, with $x^a:=\parentheses*{1/2-a,a/2,a/2,0,1/2-b/2,0,0,b/2}$, $x^b:=\parentheses*{1/2-a/2,0,0,a/2,1/2-b,b/2,b/2,0}$ and $\alpha\in\brackets*{0,1}$, is a best reply to $f^*$. If remains to show that $f^*$ is a best reply to $\phi^*$ for a suitable $\alpha$. To this end, note that from linearity of expectation and the payoff function in the values of $f$, necessarily there exists a best reply $f$ to $\phi^*$ getting values in $\braces*{0,1}$. Take $\alpha=\frac{b}{a+b}$.

Consider the following cases:
\begin{itemize}
    \item If $2a\leq b$ -- by inspection, one can check that $f\parentheses*{0}=f\parentheses*{1/2}=0$, $f\parentheses*{1}=1$ is a best reply to $\phi^*$. Moreover, it yields expected payoff of $0 \cdot \frac{b}{a+b}+\frac{2b+a}{2}\cdot \frac{a}{a+b}=\frac{a^2+2ab}{2\parentheses*{a+b}}=\reg\parentheses*{f^*,\phi^*}$. Hence, $f^*$ is also a best reply to $\phi^*$, as desired.
    \item Otherwise, a straightforward check implies that $f\parentheses*{0}=f\parentheses*{1/2}=0$, $f\parentheses*{1}=1$ is a best reply to $\phi^*$, yielding expected payoff of $\frac{2a-b}{2}\cdot \frac{b}{a+b}+\frac{2b+a}{2}\cdot \frac{a}{a+b}=\frac{a^2+4ab-b^2}{2\parentheses*{a+b}}=\reg\parentheses*{f^*,\phi^*}$; thus, $f^*$ is also a best reply to $\phi^*$, as needed.
\end{itemize}

\paragraph{{\bf Case 2: $1/2\leq a<b$.}} This case is symmetric to the previous one, with $a$ replaced by $1-b$, and $b$ replaced by $1-a$.

\paragraph{{\bf Case 3: $a\leq 1/2\leq b$.}} We argue similarly to Case~1. The definition of $h$ stays the same, but this time the constraints on $p,q$ are $p\in\brackets*{0,a/2}$ and $q\in\brackets*{b-1/2,b/2}$. By analogous arguments to Case~1, we get that for the optimal $f$ it holds that $h\parentheses*{a/2,b-1/2}=h\parentheses*{0,b/2}$. That is, $1/2-a/2+1-b+\max\braces*{a/2,b-1/2}-\parentheses*{1/2-a/2}-2\parentheses*{1/2-b/2}f\parentheses*{1/2}-b+1/2=\max\braces*{1/2-a,1/2-b/2}+a-1/2+a f\parentheses*{1/2}$. A straightforward computation shows that the value of $f\parentheses*{1/2}$ is as in the proposition statement, with the adversary's best reply being $\alpha\delta_{x^a}+\parentheses*{1-\alpha}\delta_{x^b}$, s.t.~$\alpha=\frac{b}{a+b}$, $x^a:=\parentheses*{1/2-a,a/2,a/2,0,1/2-b/2,0,0,b/2}$ and $x^b:=\parentheses*{1/2-a/2,0,0,a/2,0,1/2-b/2,1/2-b/2,b-1/2}$.
\end{proof}
\section{Proof of Theorem~\ref{thm:arb}}
\label{app:arb}
Similarly to Theorem~\ref{thm:dictator} proof, we start with a lemma lower-bounding $\reg\parentheses*{f}$.
\begin{lemma}
\label{lem:lower_arb}
Fix $f$. Let $l$ be the line through $\parentheses*{0,f\parentheses*{0}}$ and $\parentheses*{1,f\parentheses*{1}}$. Suppose there exist distinct $x_1,x_2,x_3,x_4\in\braces*{0,\frac{1}{n},\ldots,1}$ s.t.~$x_1\leq a_{\omega'}\leq x_2$ for every $\omega'\in\Omega^L$ and $x_3\leq b\leq x_4$ for every $\omega'\in\Omega^H$. Suppose further that $\parentheses*{x_1,f\parentheses*{x_1}},\parentheses*{x_2,f\parentheses*{x_2}}$ is not below $l$ and $\parentheses*{x_3,f\parentheses*{x_3}},\parentheses*{x_4,f\parentheses*{x_4}}$ not above $l$. Then:
\begin{align*}
    &\reg\parentheses*{f}\geq \reg\parentheses*{l}=\sum_{\omega'\in\Omega^H} \mu_{\omega'}u\parentheses*{\omega'} - \sum_{\omega'\in\Omega}\mu_{\omega'}\parentheses*{a_{\omega'}\parentheses*{f\parentheses*{1}-f\parentheses*{0}}+f\parentheses*{0}}u\parentheses*{\omega'}.
\end{align*}
\end{lemma}
\begin{proof}
Assume that there exist $x_1,x_2,x_3,x_4$ as described. As in Lemma~\ref{lem:lower} proof, we get that the Bayesian DM's expected utility under $l$ is independent of the information structure $\pi$; moreover, the Bayesian DM's expected utility equals its maximum possible value $\sum_{\omega'\in\Omega^H} \mu_{\omega'}u\parentheses*{\omega'}$ for $\pi$ inducing the distributions $\parentheses*{\hat{\pi}_{\omega'}}_{\omega'\in\Omega}$ over values of $\nu$ with respective expectations equal to $a_{\omega'}$ such that $\supp\parentheses*{\hat{\pi}_{\omega'}}=\braces*{x_1,x_2}$ for $\omega'\in\Omega^L$ and $\supp\parentheses*{\hat{\pi}_{\omega'}}=\braces*{x_3,x_4}$ for $\omega'\in\Omega^H$ (note that the distributions $\hat{\pi}_{\omega'}$ are uniquely determined). Thus, it is enough to prove that for the above $\pi$, the expected ignorant DM's utility under $f$ is at most as under $l$.

Indeed, by the assumption on $x_1,x_2,x_3,x_4$, the ignorant DM's utility loss in some $\omega'\in\Omega^L$ for taking the optional action compared to taking the default action is at least $a_{\omega'}\parentheses*{f\parentheses*{1}-f\parentheses*{0}}-f\parentheses*{0}\absolute*{u\parentheses*{\omega'}}$, while her utility gain in some $\omega'\in\Omega^H$ for taking the optional action compared to the default action is at most $\parentheses*{a_{\omega'}\parentheses*{f\parentheses*{1}-f\parentheses*{0}}+f\parentheses*{0}}u\parentheses*{\omega'}$. Therefore, the expected utility of the ignorant DM under $f$ is at most\\$\sum_{\omega'\in\Omega}\mu_{\omega'}\parentheses*{a_{\omega'}\parentheses*{f\parentheses*{1}-f\parentheses*{0}}+f\parentheses*{0}}u\parentheses*{\omega'}$ -- i.e., her expected utility under $l$.
\end{proof}
\begin{proof}[Proof of Theorem~\ref{thm:arb}]
Let $l$ be the line passing through $\parentheses*{0,f\parentheses*{0}}$ and $\parentheses*{1,f\parentheses*{1}}$, and let $f\parentheses*{x}\equiv mx+k$ (for constants $m,k$) be its equation. The proof that for any choice of $f$, the inequality ensured by Lemma~\ref{lem:lower_arb} holds is done exactly as in Theorem~\ref{thm:dictator} proof, with $a$ and $b$ replaced by the sets of $a_{\omega'}$ for which $\omega'\in \Omega^L$ and $\omega'\in \Omega^H$, respectively. In particular, there exists an optimal $f$ which is linear. Note that $k=f\parentheses*{0}$ and $m=f\parentheses*{1}-f\parentheses*{0}$. From Lemma~\ref{lem:lower_arb}, for any $f$:
\begin{align*}
    &\reg\parentheses*{f}\geq\sum_{\omega'\in\Omega^H} \mu_{\omega'}u\parentheses*{\omega'}-m\sum_{\omega'\in\Omega} \mu_{\omega'}a_{\omega'}u\parentheses*{\omega'}-k\sum_{\omega'\in\Omega} \mu_{\omega'}u\parentheses*{\omega'}.
\end{align*}
Moreover, for a linear $f$ we have equality. The only constraints on $m,k$ are $k=f\parentheses*{0}\in \brackets*{0,1}$ and $m+k=f\parentheses*{1}\in \brackets*{0,1}$. To minimize the bound, we must either have $k=1$ and $m\in\braces*{-1,0}$, or $k=0$ and $m\in\braces*{0,1}$. Therefore, for any $f$:
\begin{align*}
    &\reg\parentheses*{f}\geq\sum_{\omega'\in\Omega^H} \mu_{\omega'}u\parentheses*{\omega'}-\max\braces*{\sum_{\omega'\in\Omega} \mu_{\omega'} a_{\omega'} u\parentheses*{\omega'},\sum_{\omega'\in\Omega} \mu_{\omega'} \parentheses*{1-a_{\omega'}} u\parentheses*{\omega'}, \sum_{\omega'\in\Omega} \mu_{\omega'} u\parentheses*{\omega'},0}.
\end{align*}
Note that $\sum_{\omega'\in\Omega} \mu_{\omega'} a_{\omega'} u\parentheses*{\omega'}$ is the expected utility of a DM in a setting with just a single agent. As for any posterior, an agent recommends the weakly better action, this expected utility is greater or equal to the utility of any fixed action. That is: $\sum_{\omega'\in\Omega} \mu_{\omega'} a_{\omega'} u\parentheses*{\omega'}\geq\max\braces*{\sum_{\omega'\in\Omega} \mu_{\omega'} u\parentheses*{\omega'},0}$. Moreover, we have:
\begin{align*}
    &\sum_{\omega'\in\Omega} \mu_{\omega'} a_{\omega'} u\parentheses*{\omega'}+\sum_{\omega'\in\Omega} \mu_{\omega'} \parentheses*{1-a_{\omega'}} u\parentheses*{\omega'}=\sum_{\omega'\in\Omega} \mu_{\omega'} u\parentheses*{\omega'}<\sum_{\omega'\in\Omega} \mu_{\omega'} a_{\omega'} u\parentheses*{\omega'},
\end{align*}
which implies $\sum_{\omega'\in\Omega} \mu_{\omega'} \parentheses*{1-a_{\omega'}} u\parentheses*{\omega'}<0$. Thus, for any $f$: $\reg\parentheses*{f}\geq\sum_{\omega'\in\Omega^H} \mu_{\omega'}u\parentheses*{\omega'}-\sum_{\omega'\in\Omega} \mu_{\omega'} a_{\omega'} u\parentheses*{\omega'}$. Furthermore, for $f\parentheses*{x}\equiv x$ equality holds. Therefore, the random dictator rule is optimal. The uniqueness follows from the same arguments as in Theorem~\ref{thm:dictator} proof, with Lemma~\ref{lem:lower} replaced by Lemma~\ref{lem:lower_arb}, which can be done since $\Omega^L=\braces*{\omega'\in\Omega: u\parentheses*{\omega'}< 0}$, $\Omega^H=\braces*{\omega'\in\Omega: u\parentheses*{\omega'}\geq 0}$ and there exists $\omega'\in\Omega^H$ s.t.~a strict inequality~holds.
\end{proof}

\end{document}